\documentclass[11pt,preprint]{aastex}
\usepackage{amssymb,amsmath,mathrsfs,color}
\usepackage{bm}
\usepackage{hyperref}
\usepackage{multirow}

\hypersetup{pdftitle={unWISE: unblurred coadds of the WISE imaging},
pdfauthor={Dustin Lang},pdfsubject={Astronomical image processing}}

\newcommand{\arxiv}[1]{\href{http://arxiv.org/abs/#1}{arXiv:#1}}

\newcommand{\foreign}[1]{\emph{#1}}
\newcommand{\etal}{\foreign{et\,al.}}

\newcommand{\figref}[1]{\figurename~\ref{#1}}
\newcommand{\Figref}[1]{\figref{#1}}
\newcommand{\secref}[1]{section~\ref{#1}}
\newcommand{\thou}{,\!000}

\newcommand{\niceurl}[1]{\href{#1}{\textsl{#1}}}

\newcommand{\fimg}[1]{#1}

\newcommand{\img}{\fimg{I}}
\newcommand{\wt}{w}
\newcommand{\mask}{\fimg{M}}
\newcommand{\good}{\fimg{G}}
\newcommand{\covar}{\fimg{V}}
\newcommand{\coacc}{\fimg{K}}
\newcommand{\coadd}{\fimg{C}}
\newcommand{\cowt}{\fimg{W}}
\newcommand{\ppchi}{\fimg{\chi}}
\newcommand{\con}{\fimg{N}}
\newcommand{\copp}{\fimg{S}}
\newcommand{\sgood}{\fimg{H}}
\newcommand{\simg}{\fimg{J}}
\newcommand{\sigprior}{\fimg{P}}
\newcommand{\nprior}{\nu}
\newcommand{\fluxsys}{e}
\newcommand{\patch}[2]{\textrm{patch}(#1 \, , \, #2)}
\newcommand{\binaryand}{\wedge}
\newcommand{\abs}[1]{\lvert#1\rvert}

\newcommand{\bwfig}[1]{#1}

\begin{document}

\title{unWISE: unblurred coadds of the WISE imaging}
%\author{Dustin~Lang\altaffilmark{1}}
\author{Dustin~Lang}
\email{dstn@cmu.edu}
%\altaffiltext{1}
\affil%
{McWilliams Center for Cosmology,
  Department of Physics, \\ Carnegie Mellon University,
  5000 Forbes Ave, Pittsburgh, PA 15213, USA}
\date{}
\shorttitle{unWISE coadds}
\shortauthors{Lang}

\begin{abstract}
The Wide-Field Infrared Survey Explorer (WISE; \citealt{wright})
satellite observed the full sky in four mid-infrared bands in the
$2.8$ to $28$ $\mu m$ range.  The primary mission was completed in
2010.  The WISE team have done a superb job of producing a series of
high-quality, well-documented, complete Data Releases in a timely
manner.  However, the ``Atlas Image'' coadds that are part of the
recent AllWISE and previous data releases were intentionally blurred.
Convolving the images by the point-spread function while coadding
results in ``matched-filtered'' images that are close to optimal for
detecting isolated point sources.  But these matched-filtered images
are sub-optimal or inappropriate for other purposes.  For
example, we are photometering the WISE images at the locations of
sources detected in the Sloan Digital Sky Survey \citep{york} through
forward modeling, and this blurring decreases the available
signal-to-noise by effectively broadening the point-spread function.
This paper presents a new set of coadds of the WISE images that have not been
blurred.  These images retain the intrinsic resolution of the data and
are appropriate for photometry preserving the available
signal-to-noise.
Users should be cautioned, however, that the W3- and W4-band coadds contain
artifacts around large, bright structures (large galaxies, dusty
nebulae, etc); eliminating these artifacts is the subject of ongoing
work.
These new coadds, and the code used to produce them, are publicly
available at \niceurl{http://unwise.me}.
\end{abstract}

\section{Introduction}

\begin{figure}[t!]
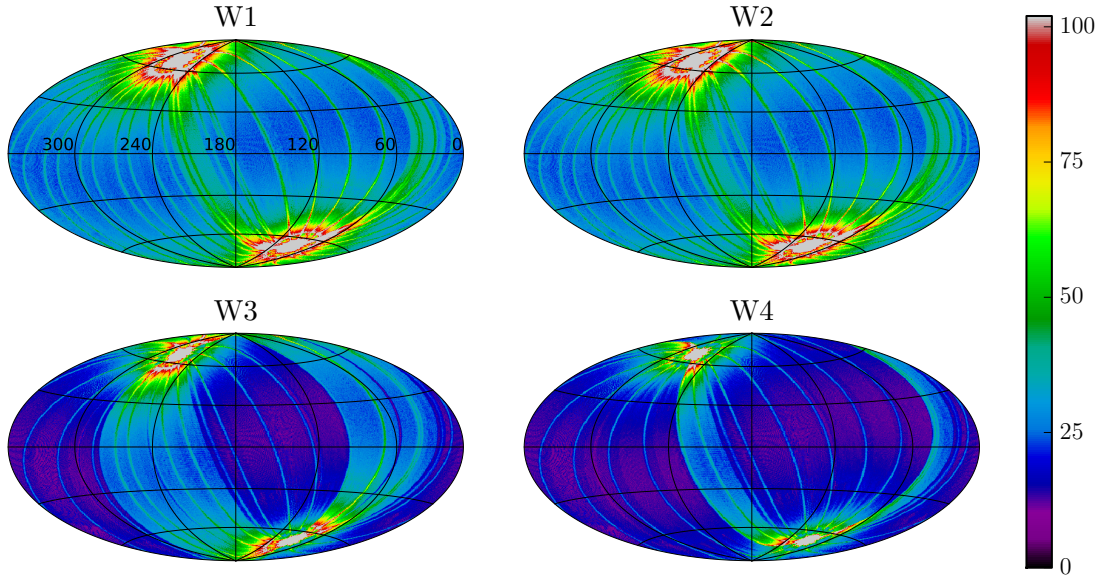

\begin{center}
\newlength{\figw}
\setlength{\figw}{0.39\textwidth}
\begin{tabular}{@{}ccc@{}}
W1 & W2 &
\multirow{3}{*}{\includegraphics[height=1.2\figw]{\bwfig{plots6/cov-00}}} \\
\includegraphics[width=\figw]{\bwfig{plots6/cov-01}} &
\includegraphics[width=\figw]{\bwfig{plots6/cov-02}} &
\\
W3 & W4 & \\
\includegraphics[width=\figw]{\bwfig{plots6/cov-03}} &
\includegraphics[width=\figw]{\bwfig{plots6/cov-04}} &
\end{tabular}
\end{center}
\caption{WISE coverage: number of exposures in W1, W2, W3, and W4
  bands including all three mission phases.  The projection is a
  Hammer--Aitoff all-sky projection in RA,Dec with RA=180, Dec=0 in the
  center.  The extremely high-coverage areas are the Ecliptic poles.
  \label{fig:coverage}
}
\end{figure}

The Wide-Field Infrared Survey Explorer (WISE) mission is described in
detail by \citet{wright}; I only briefly summarize it here.  The
instrument measures four infrared bands centered on 3.4 $\mu m$ (W1),
4.6 $\mu m$ (W2), 12 $\mu m$ (W3), and 22 $\mu m$ (W4), using HgCdTe
arrays for the shorter W1 and W2 bands, and Si:As for W3 and W4.
Survey operations began 2010 Jan 14.  Coverage of the full sky was
completed 2010 Jul 17.  Observations continued until 2010 Aug 6, when
the solid hydrogen cryogen ran out and the longest band, W4, became
unusable.  ``Three-band'' observations using the remaining W1, W2, and
W3 channels continued until 2010 Sep 29, and the ``NEOWISE post-cryo''
program \citep{mainzer} continued observations in the shortest W1 and
W2 bands until 2011 Feb 1.  The WISE telescope is 40 cm and provides a
field of view 47 by 47 arcmin which is imaged simultaneously via
dichroics on the four detectors.  Exposure time was 7.7 sec (W1 and
W2) and 8.8 sec (W3 and W4).  The sky was scanned from Ecliptic pole
to pole, with roughly 90\% overlap from scan to scan.  Over 99\% of
the sky has 11 or more exposures in W3 and W4, and 23 or more
exposures in W1 and W2.  Median coverage is 33 exposures in W1 and W2,
24 in W3, and 16 in W4.  Coverage increases toward the Ecliptic poles,
with over $3\thou$ exposures in each band at the poles.  See
\figref{fig:coverage}.

%   exposures:
%                   W 1    W 2    W3    W4
%   min:              20    20     8     8
%   percentile 1 :    23    23    11    11
%   percentile 2 :    23    23    11    11
%   percentile 5 :    24    24    12    12
%   percentile 10 :   25    25    13    12
%   percentile 50 :   33    33    24    16
%   percentile 90 :   60    60    43    35
%   percentile 95 :   80    80    56    45
%   percentile 98 :  106   107    79    61
%   percentile 99 :  129   129    98    78
%   max:            5885  5883  4145  3276
%
% From counting healpixes in WCSes (nside=200)

%
% These plots are from check-coadd.py: composites() function
% rev 24153
%
\begin{figure}[t!]
\begin{center}
\begin{tabular}{@{}c@{\hspace{0.03\textwidth}}c@{}}
WISE & unWISE \\
\includegraphics[width=0.4\textwidth]{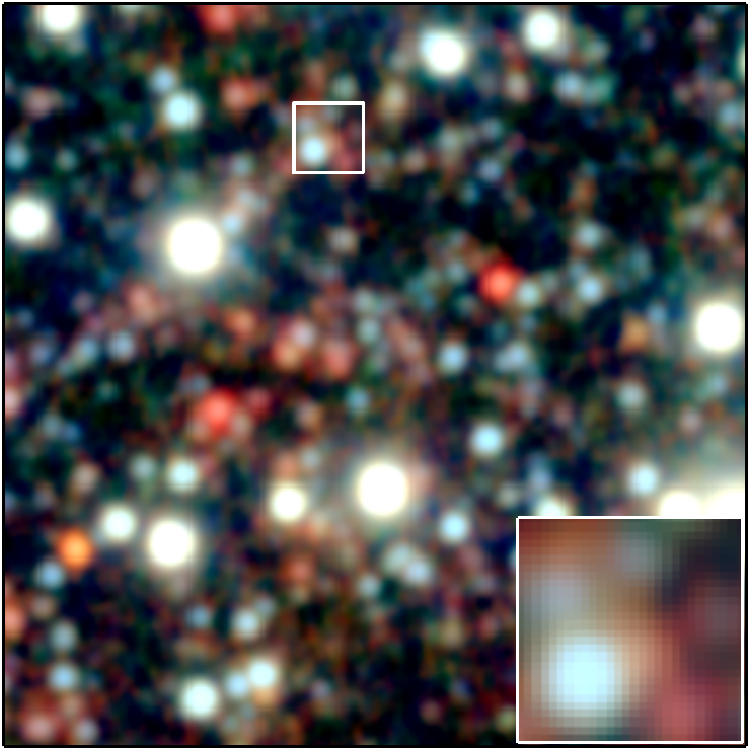} &
\includegraphics[width=0.4\textwidth]{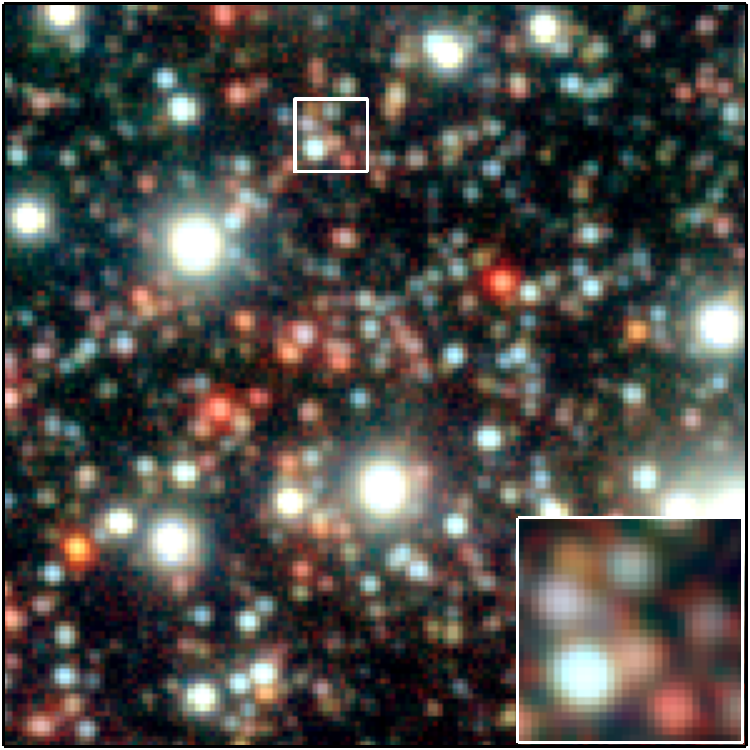}
\end{tabular}
\end{center}
\caption{An example image: the coadd tile containing the north
  Ecliptic pole (2709p666), three-band composites (W1,W2,W3).  These
  subimages are roughly $7\times7$ arcminutes.  \emph{Left:} AllWISE
  Release Atlas Image.  \emph{Right:} The unWISE coadd.  The insets
  show a zoom-in on a region containing crowded sources.  In the WISE
  Atlas Image they are blended together, while in the unWISE coadd
  they are clearly distinct.  These composite images are made as per
  \citet{lupton}.
  \label{fig:npole}}
\end{figure}

%
% These are from check-coadd.py at rev 24020 (coverage_plots() function)
% B/W: rev 24630
%

The pixel scale is roughly 2.75 arcseconds per pixel, and the
delivered images are 1016 by 1016 pixels.  The W4 images were binned
2-by-2 onboard.  The 4-band primary mission collected $\sim$1.5
million four-band image sets.  The 3-band phase collected
$\sim400\thou$ three-band image sets, and the NEOWISE post-cryo phase
collected $\sim900\thou$ two-band image sets.

The AllWISE Data Release\footnote{Explanatory Supplement to the
  AllWISE Data Release Products,
  \niceurl{http://wise2.ipac.caltech.edu/docs/release/allwise/expsup/}}
was announced 2013 Nov 13, and includes data taken during all three
mission phases.  The data products in the AllWISE Release include a
source catalog of nearly 750 million sources, a database of photometry
in the individual frames at each source position, and ``Atlas
Images'': coadded matched-filtered images.  The previous All-Sky Data
Release\footnote{Explanatory Supplement to the All-Sky Data Release
  Products,
  \niceurl{http://wise2.ipac.caltech.edu/docs/release/allsky/expsup/}}
also included calibration individual exposures (``level 1b'' images),
which I use here.  The algorithms are described in considerable depth
in the Explanatory Supplement documents, and the Atlas Image coadds
are described by \cite{awaic}.  I review briefly relevant parts here.

The Atlas Images in the AllWISE Release are intentionally convolved by
a model of the point-spread function, producing what are sometimes
called ``detection maps'' because they are close to optimal for
detecting isolated point sources.  This operation, however, blurs the
images, making them less than ideal for some purposes.  In this work,
I produce coadds that do not include this blurring step; see
\figref{fig:npole} for an example.  As a result, my coadds maintain
the full resolution of the original images and are appropriate for
making photometric measurements.  While in principle I typically
advocate making photometric measurements by simultaneously
forward-modeling the individual exposures, the WISE instrument is
remarkably stable and has a nearly isotropic point-spread function, so
creating a coadd results in little loss of information about the
non-variable sky.

The WISE Atlas Images (and my coadds) start with the calibrated
``level 1b'' individual exposures.  These have been calibrated
astrometrically and photometrically, flat-fielded, and had static
artifacts (bad pixels, etc) flagged.  The first step is selecting the
frames to be included in the coadd.  The criteria used (which I also
follow) are to omit W3 and W4 frames within 2000 seconds of an
annealing event (the W3 and W4 Si:As arrays were periodically
heated---``annealed''---to remove latent images); frames with a bad
frame-level quality-assurance score from the WISE image-processing
pipeline; and frames taken during some on-orbit experiments.  For each
frame, the photometric calibration is applied to convert the images to
common units.  Next, background level matching is performed by
matching the median of a new image to that of the existing coadd, in
the pixels where they overlap.  Frames suspected of being affected by
moon-glow are subjected to tests on their pixel distributions; a
rapidly-varying background due to moon glow increases the pixel RMS
significantly above that of an image not affected by the moon.
Downstream, the coadding process uses median statistics, so is quite
robust to outliers; some moon-contaminated frames can be tolerated,
but moon-contaminated frames are rejected if they form too large a
fraction of the frames.  Next, all frames to be included in the
coadd are projected and interpolated onto the coadd frame in order to
detect outliers.  A ``pixel overlap-area weighting method,'' or
top-hat interpolation kernel, is used in this step.  The list of
interpolated pixel values for each coadd pixel is stored so that
robust statistics (the median and approximate median-of-absolute
differences) can be computed pixelwise.  A set of heuristics is used
to flag outlier pixels using the computed robust statistics.  An
additional set of heuristics expands masked regions so that regions
containing several masked pixel become completely masked.  Finally,
after outlier masking, the final interpolation is performed.  This
uses an approximation to ``pixel response function'' weighting,
equivalent to first smoothing the image by its point-spread function
and then resampling that image.  In practice, a sub-pixelized PRF
representation and nearest-neighbor interpolation are used.

% Created by boxes.py at rev 24022
\begin{figure}
%\newlength{figw}
%\setlength{figw}{0.24\textwidth}
\newlength{\figh}
\setlength{\figh}{0.18\textwidth}
\begin{center}
\begin{tabular}{@{}c@{}c@{}c@{}c@{}}
Top-hat & Fine top-hat & Lanczos-3 & Lanczos-5 \\
\includegraphics[height=\figh]{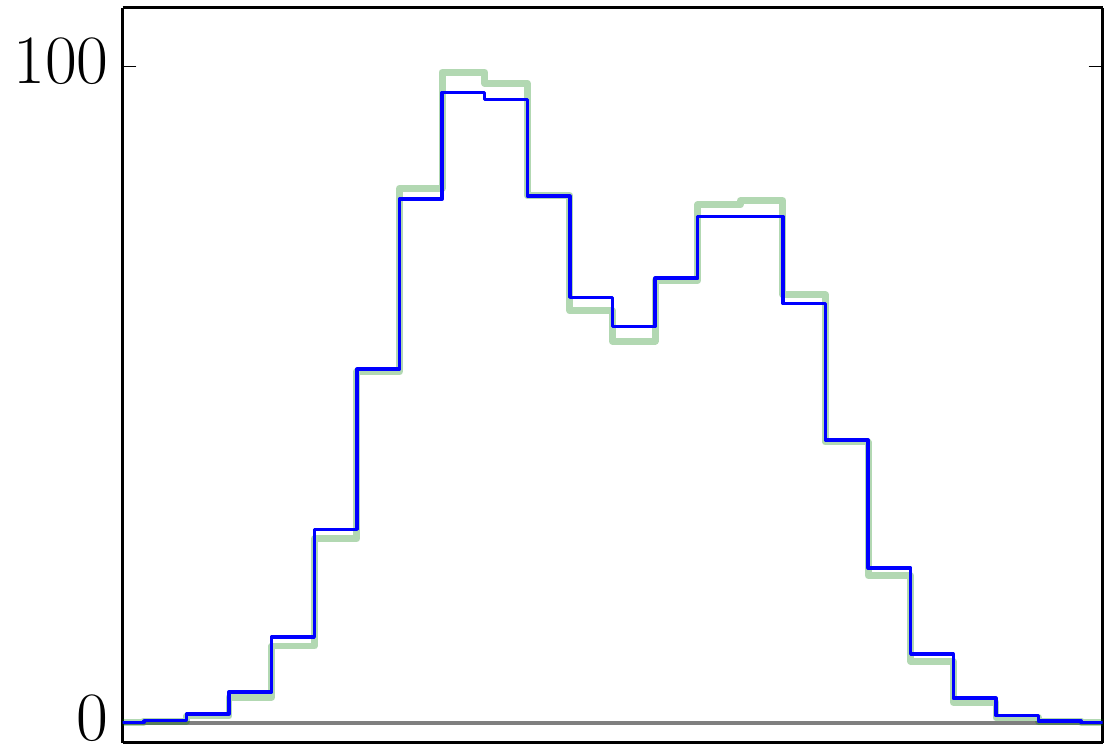} &
\includegraphics[height=\figh]{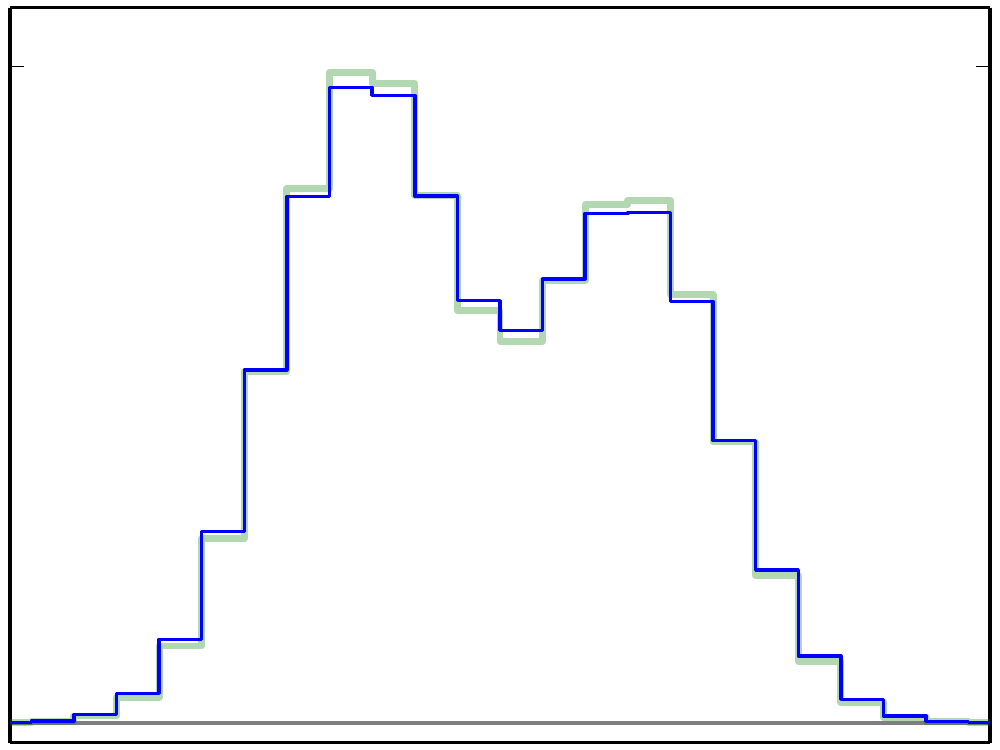} &
\includegraphics[height=\figh]{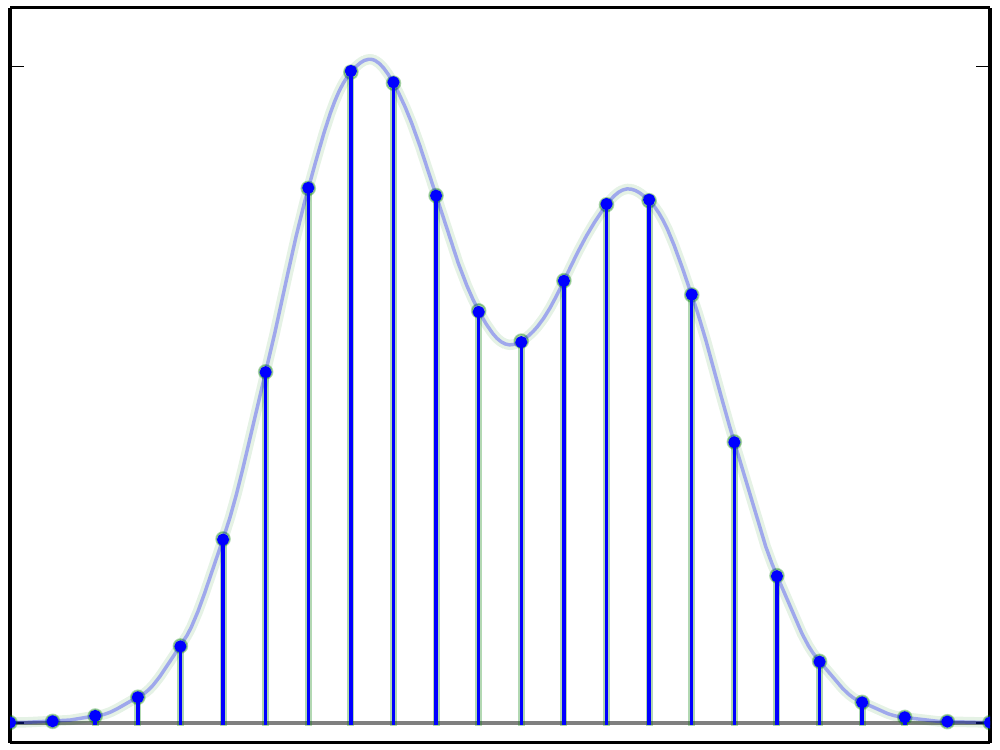} &
\includegraphics[height=\figh]{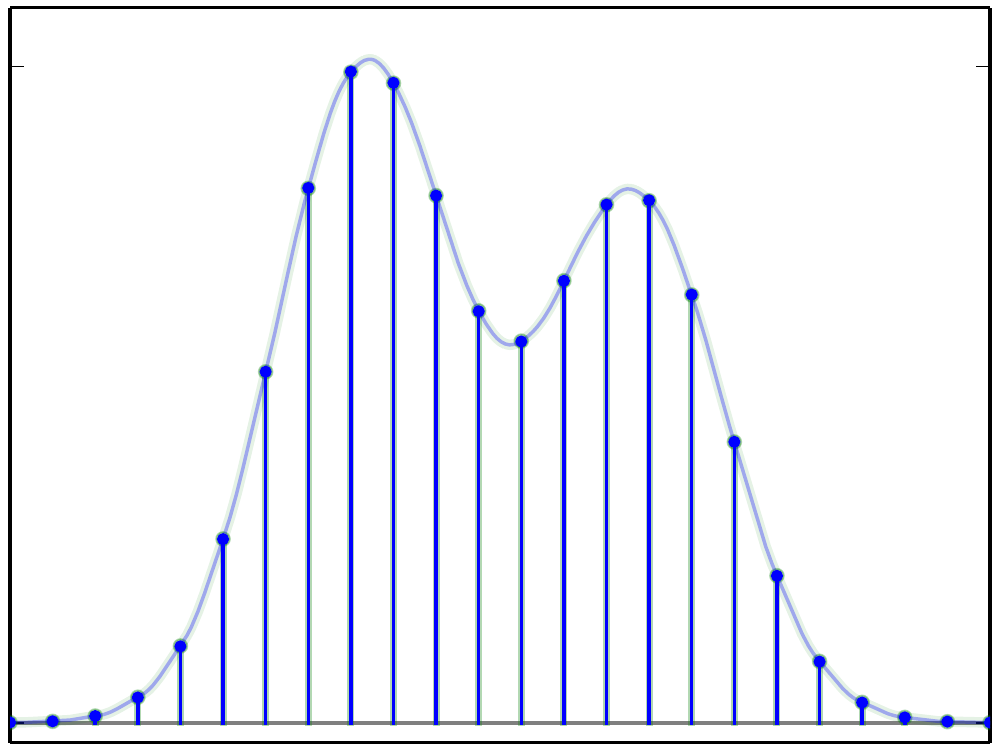} \\
\includegraphics[height=\figh]{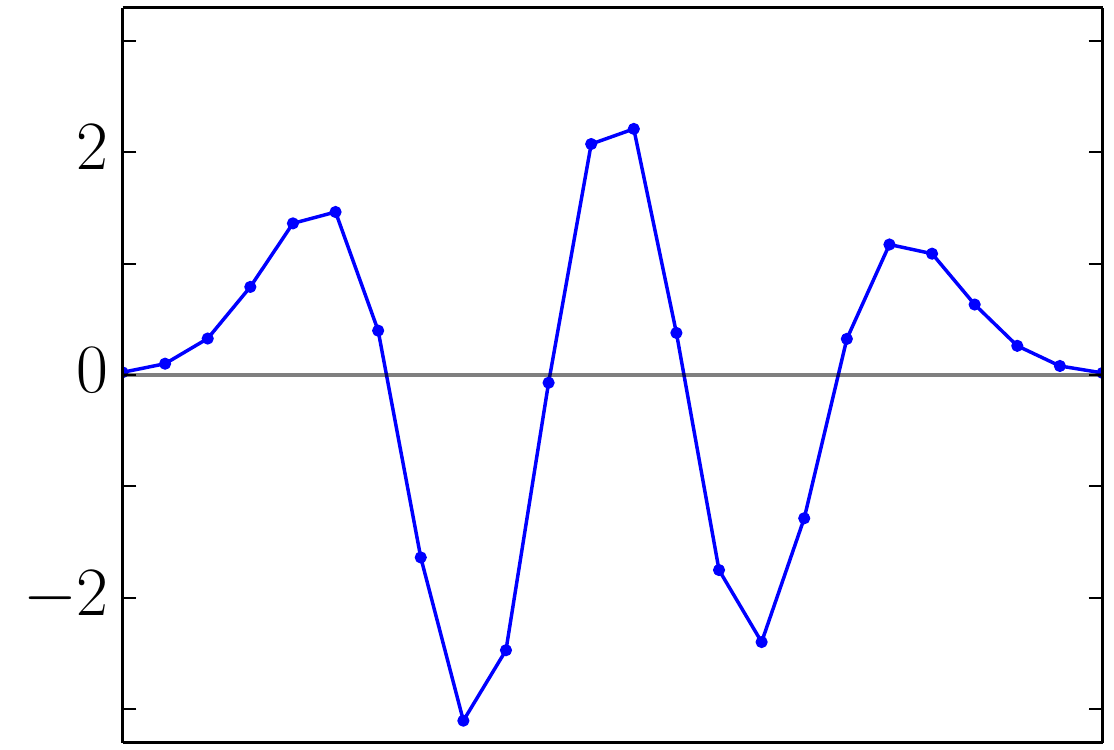} &
\includegraphics[height=\figh]{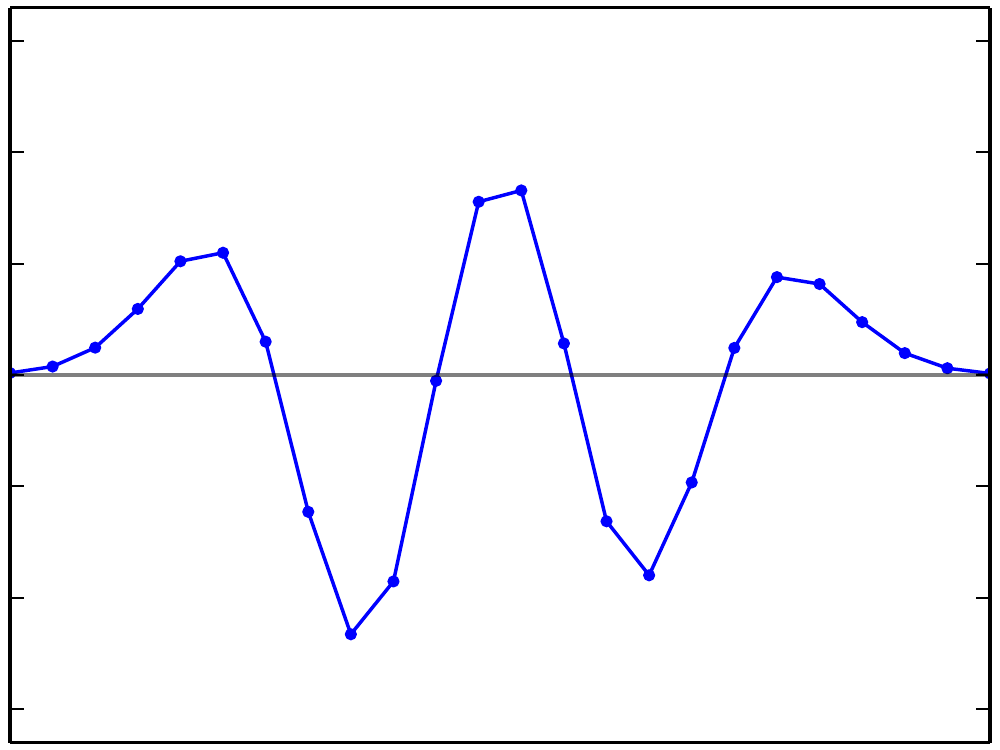} &
\includegraphics[height=\figh]{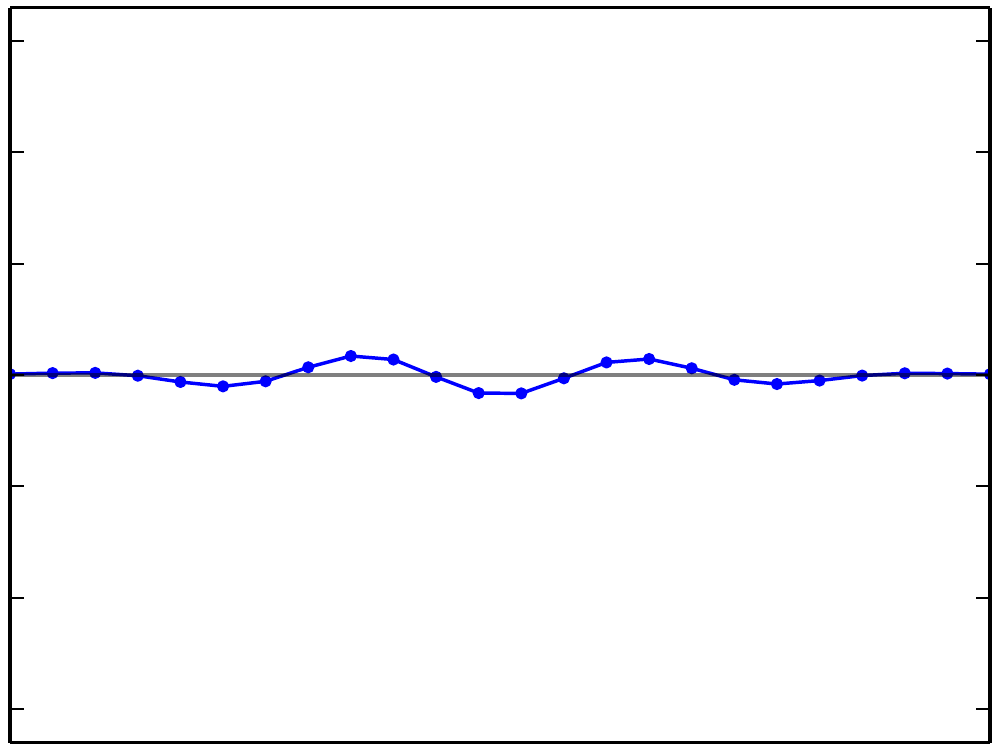} &
\includegraphics[height=\figh]{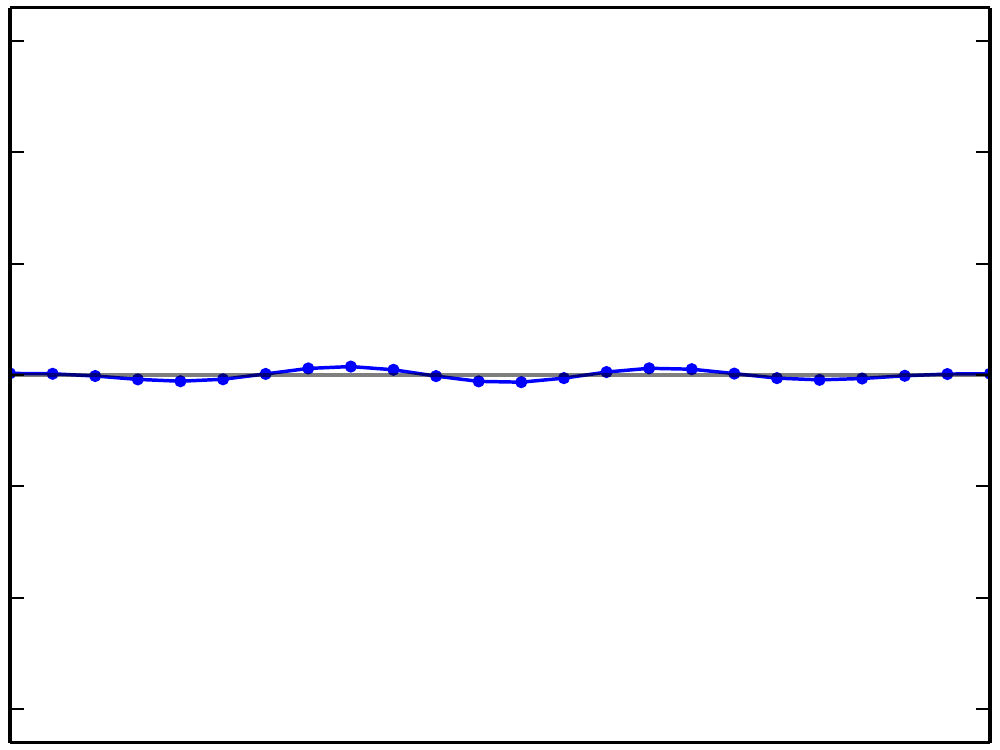} \\
\includegraphics[height=\figh]{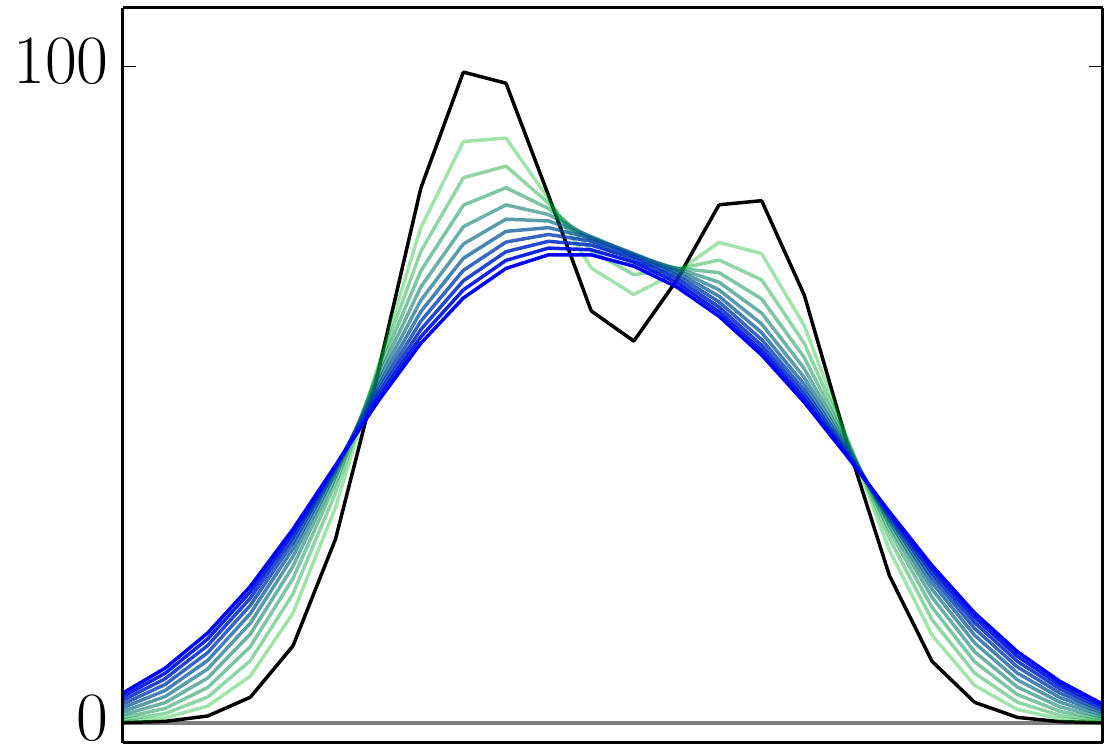} &
\includegraphics[height=\figh]{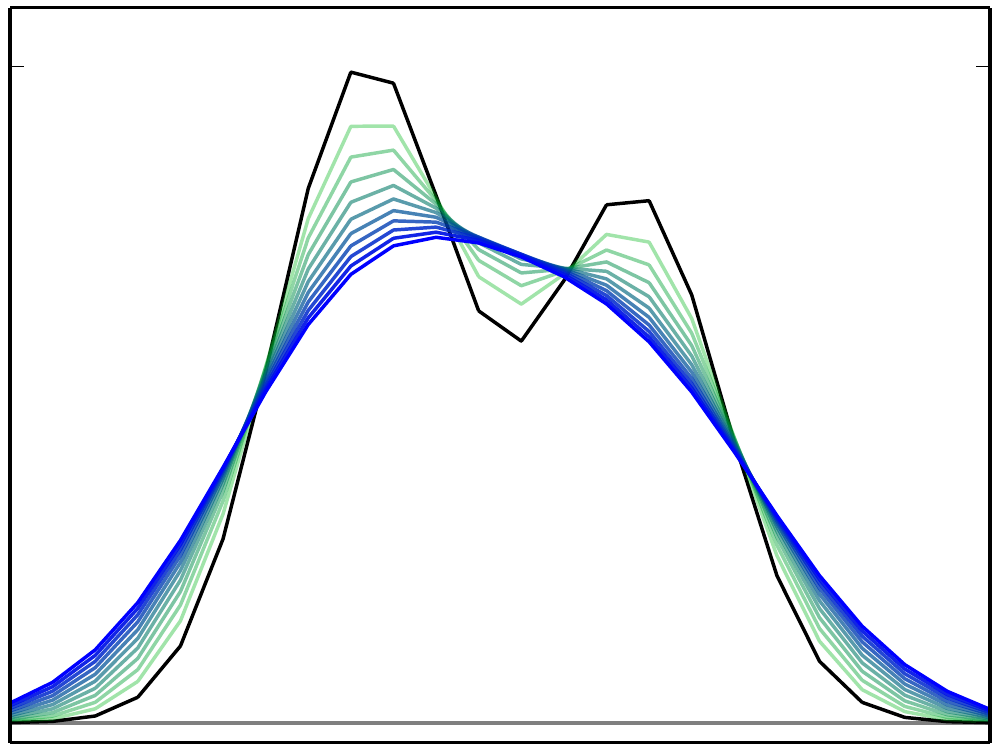} &
\includegraphics[height=\figh]{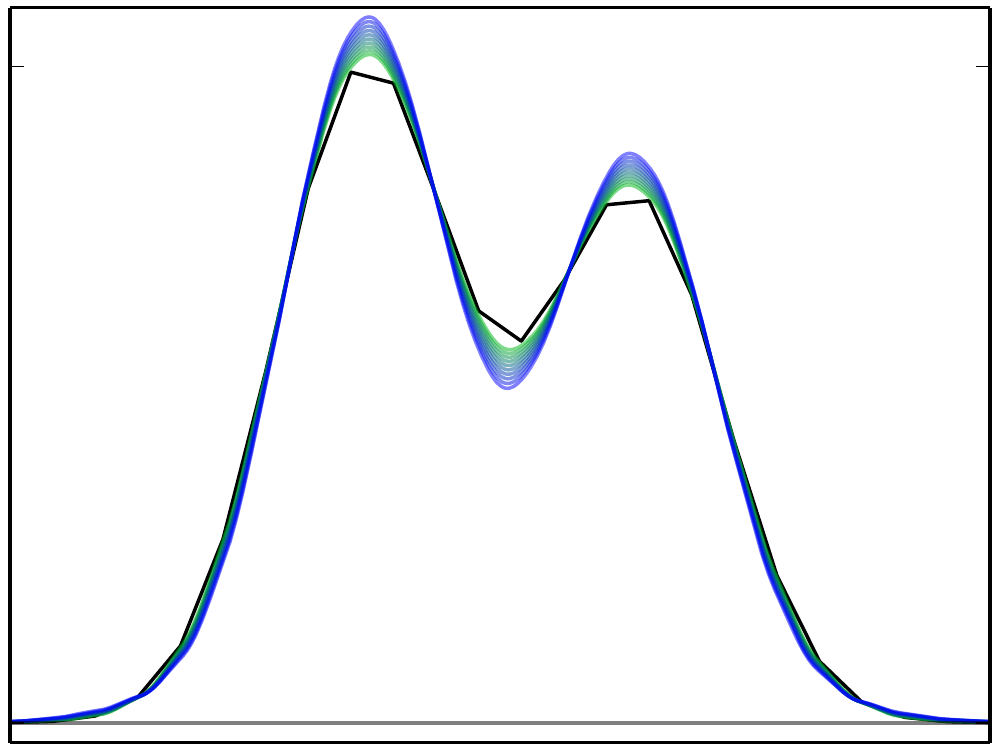} &
\includegraphics[height=\figh]{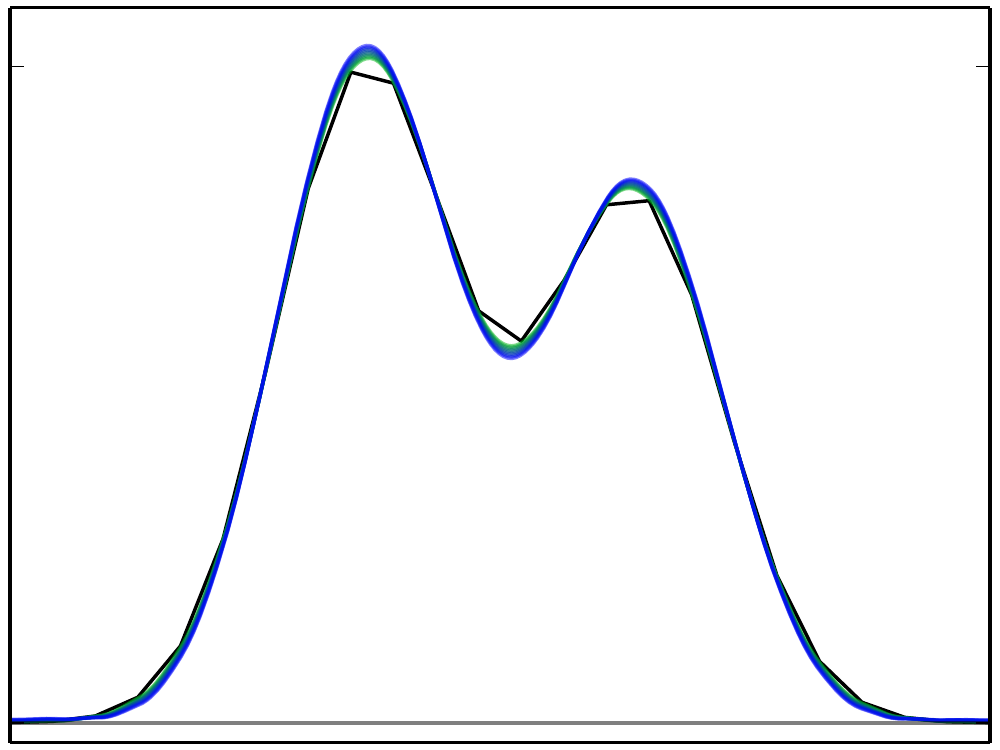}
\end{tabular}
\end{center}
\caption{\small Comparison of ``pixel overlap'' (or ``top-hat''; used
  in the AllWISE release; left) and Lanczos (used in the unWISE
  coadds; right) resampling, in one dimension.  In this demonstration,
  I shift (by resampling) the signal by 0.2 pixels, and then shift
  the resampled signal back.  In the second column (``Fine top-hat''),
  the resampled pixels are at twice the original pixel resolution (as
  done in the AllWISE Release).  \emph{Top:} A row of pixels; in the
  ``pixel overlap'' view, pixels are little boxes, while in my view
  they are samples of an underlying function.  The original signal is
  shown by the faint green line, and the result after resampling by
  0.2 pixels and then sampling back is shown in solid blue.
  \emph{Middle row:} Resampling error: difference between resampled
  and original signal. \emph{Bottom:} Repeating this process of
  resampling the signal back and forth 30 times results in marked
  smoothing of the signal in the case of pixel overlap resampling, but
  little loss of fidelity for the case of Lanczos resampling.
  The signal is a mixture of two Gaussians so is technically not
  band limited; some aliasing error is to be expected.
  \label{fig:boxes}}
\end{figure}

The primary difference between my coadds and those of the AllWISE
Release is in the interpolation kernels we use.  This belies a
difference in philosophy as well.  The WISE Explanatory Supplement
documents describe pixels as though they are ``little boxes''.  This
picture leads to confusion and heuristics with no theoretical basis in
sampling or information theory.  Unfortunately, this ``pixels are
little boxes'' view has persisted for some time in astronomical image
processing, perhaps most strikingly in the form of the ``drizzle''
process predominantly used for Hubble Space Telescope images
\citep{drizzle}.%
\footnote{Since HST images are typically undersampled, the resampling
  approach advocated here does not apply; my point is simply that
  the ``drizzle'' algorithm has no justification in sampling theory;
  for alternatives see \cite{lauer, rowe, mahmoudian}.}  In contrast,
I think of detector pixels as taking (noisy) delta-function samples
of an underlying smooth, continuous function: the pixel-convolved
point-spread function convolved with the scene.  Once the pixels have
been read out, it is not necessary or helpful to think of them as
``little boxes''; they are simply samples of an underlying
two-dimensional function.  As I show below, the WISE pixels
well-sample the point-spread function and thus this underlying
function, so the Nyquist--Shannon sampling theorem applies; a discrete
set of samples contains sufficient information to reconstruct the
original signal.

Briefly, the Nyquist--Shannon sampling theorem, in one dimension,
states that if a continuous function contains no frequency components
above a band-limit frequency $B$, then a discrete set of samples
spaced $1/2B$ apart completely characterise the function; the original
function can be reconstructed exactly through sinc interpolation of the
samples.  A good review article is \cite{jerri}.
In the context of astronomical images, an image satisfies the sampling
criterion---it is ``well-sampled''---if the underlying function
contains no spatial frequency components (Fourier modes) shorter than
2 pixels.  In that case, a discrete set of image pixels (samples) are
sufficient to completely characterise the underlying function.  As
mentioned above, the underlying function is the astronomical scene
convolved by the pixel-convolved point-spread function; since the
scene can contain (nearly) point sources, the well-sampled condition
depends on the pixel size relative to the pixel-convolved point-spread
function.
Given a well-sampled image, we can resample it to a different pixel
grid, without loss of information---as long as the target pixel grid
also well-samples the image---through sinc interpolation.  The sinc
function is the Fourier transform of the ideal low-pass filter,
leaving all frequencies below the band limit untouched and removing
all frequencies above it.  The sinc interpolation kernel has infinite
extent, so in practice we taper the sinc kernel to give it finite
extent.  In this work, I use the popular Lanczos kernel as a tapered
approximation to the sinc kernel.

\Figref{fig:boxes} shows the loss of information resulting from
using the top-hat kernel compared to the Lanczos kernel.

%
% These figures come from fourier.py, function wise_psf
% b/w at rev 24632
%
\begin{figure}
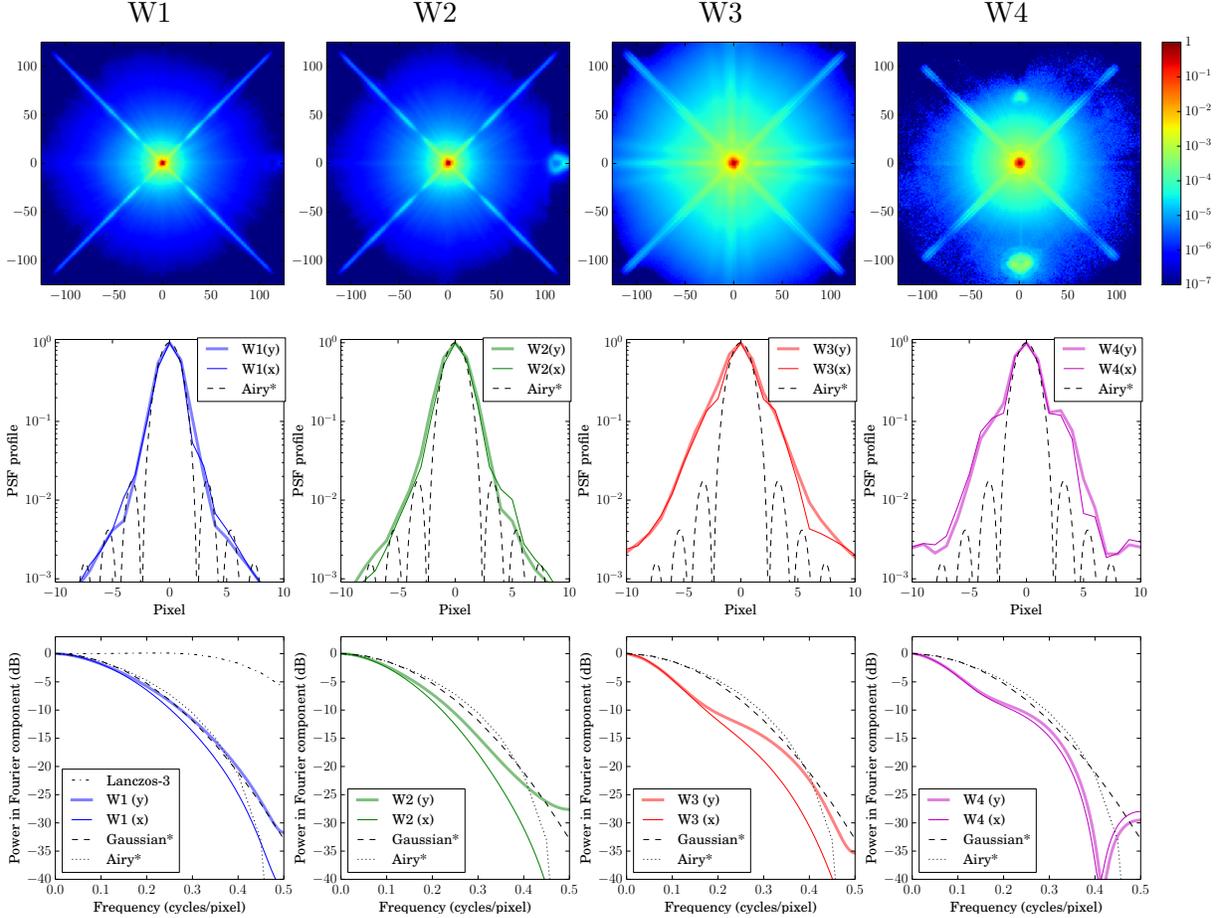

\begin{center}
\begin{tabular}{c@{}c@{}c@{}c@{\hspace{0.01\textwidth}}c}
W1 & W2 & W3 & W4 & \\
\includegraphics[height=0.23\textwidth]{\bwfig{wisepsf-01}} &
\includegraphics[height=0.23\textwidth]{\bwfig{wisepsf-02}} &
\includegraphics[height=0.23\textwidth]{\bwfig{wisepsf-03}} &
\includegraphics[height=0.23\textwidth]{\bwfig{wisepsf-04}} &
\includegraphics[height=0.23\textwidth]{\bwfig{wisepsf-00}} \\
\includegraphics[height=0.23\textwidth]{\bwfig{wisepsf-05}} &
\includegraphics[height=0.23\textwidth]{\bwfig{wisepsf-06}} &
\includegraphics[height=0.23\textwidth]{\bwfig{wisepsf-07}} &
\includegraphics[height=0.23\textwidth]{\bwfig{wisepsf-08}} & \\
\includegraphics[height=0.23\textwidth]{\bwfig{wisepsf-09}} &
\includegraphics[height=0.23\textwidth]{\bwfig{wisepsf-10}} &
\includegraphics[height=0.23\textwidth]{\bwfig{wisepsf-11}} &
\includegraphics[height=0.23\textwidth]{\bwfig{wisepsf-12}} &
\end{tabular}
\end{center}
\caption{\emph{Top row:} WISE individual-exposure PSF models from
  \cite{meisner}, at the center of the field of view, shown on a log
  scale with dynamic range $10^7$, and size $250\times250$ pixels.
  \emph{Middle row:} Slices through the PSF model in the $x$ and $y$
  directions, plus the critically-sampled Airy profile,
  $(4 \mathrm{J}_{1}(\pi x/2) / (\pi x))^2$; the PSF models
  are only a little wider.
  \emph{Bottom row:} Power spectra of 1-d slices of the PSF models
  in the $x$ and $y$ directions, along with critically-sampled Airy profile
  and a Gaussian with the same full-width at half max ($\sigma \sim 0.88$).
  The PSF models have
  a small amount of power outside the band limit, roughly comparable to the
  Gaussian.
  These PSF models are constructed from WISE data so include some noise
  which also contributes to the power spectrum.
  The leftmost (W1) panel also shows the power spectrum of the
  Lanczos-3 interpolation kernel used in this work: it is mostly flat, but
  slightly attenuates high-frequency components.
  When I measure isolated point sources in the individual exposures
  or the unWISE coadds, I find similar profiles and power spectra.
  \label{fig:psf}}
\end{figure}

\Figref{fig:psf} shows the WISE point-spread function models from
\cite{meisner} and their Fourier transforms.  These models are
constructed by combining a large number of bright stars (to get the
wings) and moderately bright stars (to get the core, which is
saturated in bright stars).  The PSF profiles roughly follow the
envelope of an Airy profile scaled to be critically sampled,
suggesting that the WISE images are approximately critically sampled.
The full-widths at half-max quoted in the WISE Explanatory Supplement
are 6.1 and 5.6 arcseconds for the major and minor axes in W1, 6.8 and
6.1 for W2, 7.4 and 6.1 for W3, and 12.0 and 11.7 for W4; for
comparison, the rule-of-thumb of $\sim 2.1$ pixels across the FWHM
(which is derived from the width of the critically-sampled Airy
profile) would be 5.7 arcseconds.
The Fourier transform of the PSF models shows a small amount of power
at the band limit, comparable to that of a Gaussian whose full-width
at half-maximum equals that of the critically-sampled Airy.  In
practice, this indicates that we can treat the WISE images as
well-sampled; if there is indeed power outside the band limit, we will
incur a small aliasing error when we resample the images.

In the case of the WISE Atlas Images, in addition to resampling to the
coadd pixel grid, there is also an extra convolution by the
point-spread function.  In order to avoid the issues with top-hat
resampling shown in \figref{fig:boxes}, the Atlas Image pixels
were chosen to have twice the resolution of the input images.  In
contrast, I do not perform any additional convolutions on the input
images, and I produce coadds at the original nominal pixel scale.
The resulting point-spread function widths are roughly $8.5$
arcseconds FWHM for the AllWISE Release Atlas Images in W1, W2, and W3
bands, versus roughly $6$ arcseconds for my coadds; the W4 PSF is
roughly twice as wide.

\section{Method}

The overall procedure follows that of the AllWISE Release, though in
general each step is simpler.  I have used the high-quality data
products provided by the WISE pipeline as far as possible.  I use the
same RA,Dec centers and axis-aligned orientations as the Atlas Image
tiles.  My coadds are $2048 \times 2048$ pixels at the nominal native
pixel scale, $2.75$ arcseconds per pixel, rather than the $4095 \times
4095$ images at $1.375$ arcseconds per pixel chosen in the AllWISE
Release.  Since the WISE images are well sampled, there is nothing to
be gained in my formulation by producing coadds at double the
resolution of the inputs.  For the sake of consistency, my W4 coadds
are on the same $2.75$ arcsecond per pixel grid as the other bands,
even though the input images have only half the resolution.

I produce a coadd in two phases.  In the first, I accumulate an
initial mean and variance estimate for the coadd pixels and use this
to mask outlier pixels in the input frames.  I mask and patch these
pixels, then produce the final coadd.

I select input frames from all three mission phases that touch the
RA,Dec box of the target coadd tile.  I remove any frames with WISE
pipeline-assigned frame-level quality score of zero.  For W3 and W4,
I remove frames within 2000 seconds of an anneal.  For W4, I remove
scans in the range 03752a to 03761b inclusive (as do the WISE team),
in which the W4 bias level was changed for testing purposes.

I drop frames that appear to be contaminated by scattered moonlight,
following the procedure used by the WISE team.  After all other cuts
have been made, I examine frames that are not within the moon mask
(ie, unlikely to be strongly affected by moon-glow).  I use the robust
estimate of the pixel standard deviation measured in the WISE pipeline
(the median minus 16\textsuperscript{th} percentile), computing the
median and median absolute difference of this quantity in the
uncontaminated frames.  This roughly characterizes the ``typical''
background variations in the coadd tile.  I reject any frame within
the moon mask whose pixel standard deviation is greater than the
median plus 5 median absolute differences times a Gaussian correction
factor of $1.4826$.

%\emph{w?intmed16ptile},

% These plots were produced via:
% python -u unwise-coadd.py --outdir e --threads 8 --plots --size 200 1000
% at rev 23617.
%
\begin{figure}
\begin{center}
\begin{tabular}{@{}ccc@{}}
Resampled image ($\img_i$) &
Mask ($\mask_i$) &
Unmasked pixels ($\good_i$) \\
\includegraphics[width=0.25\textwidth]{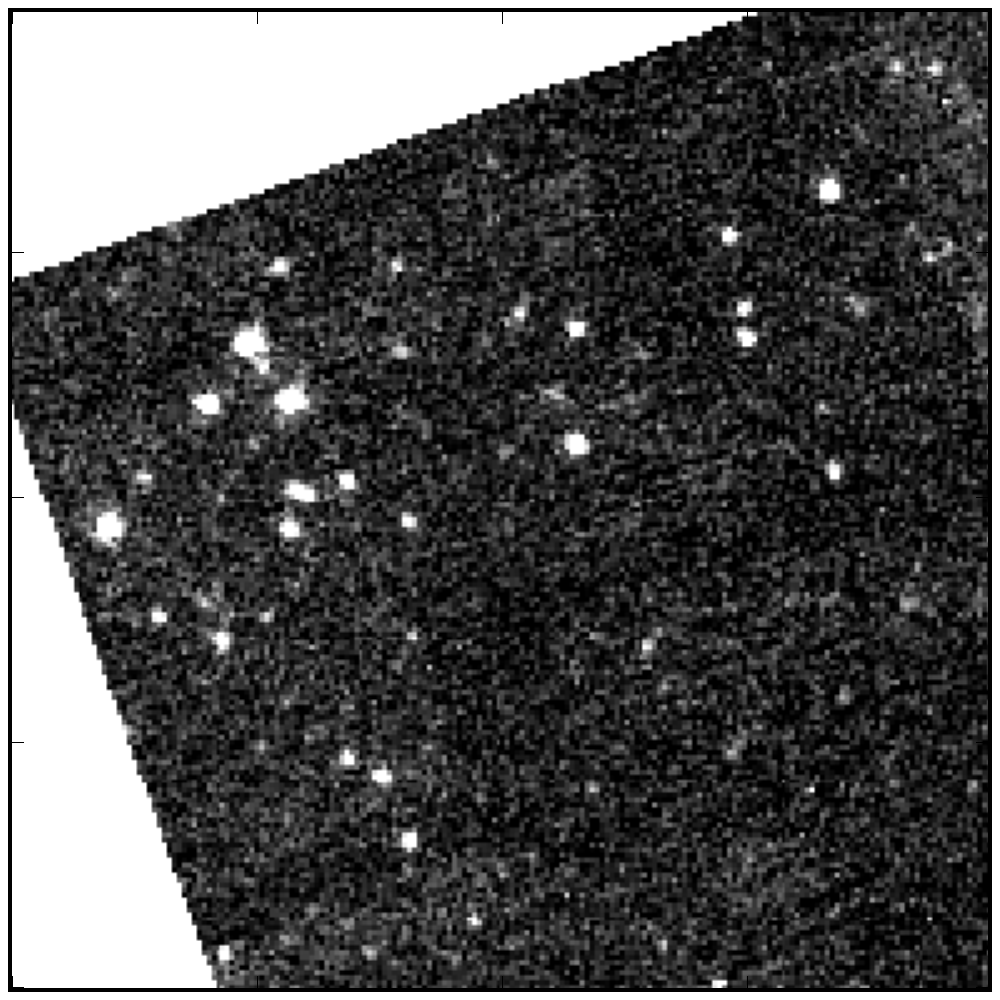} &
\includegraphics[width=0.25\textwidth]{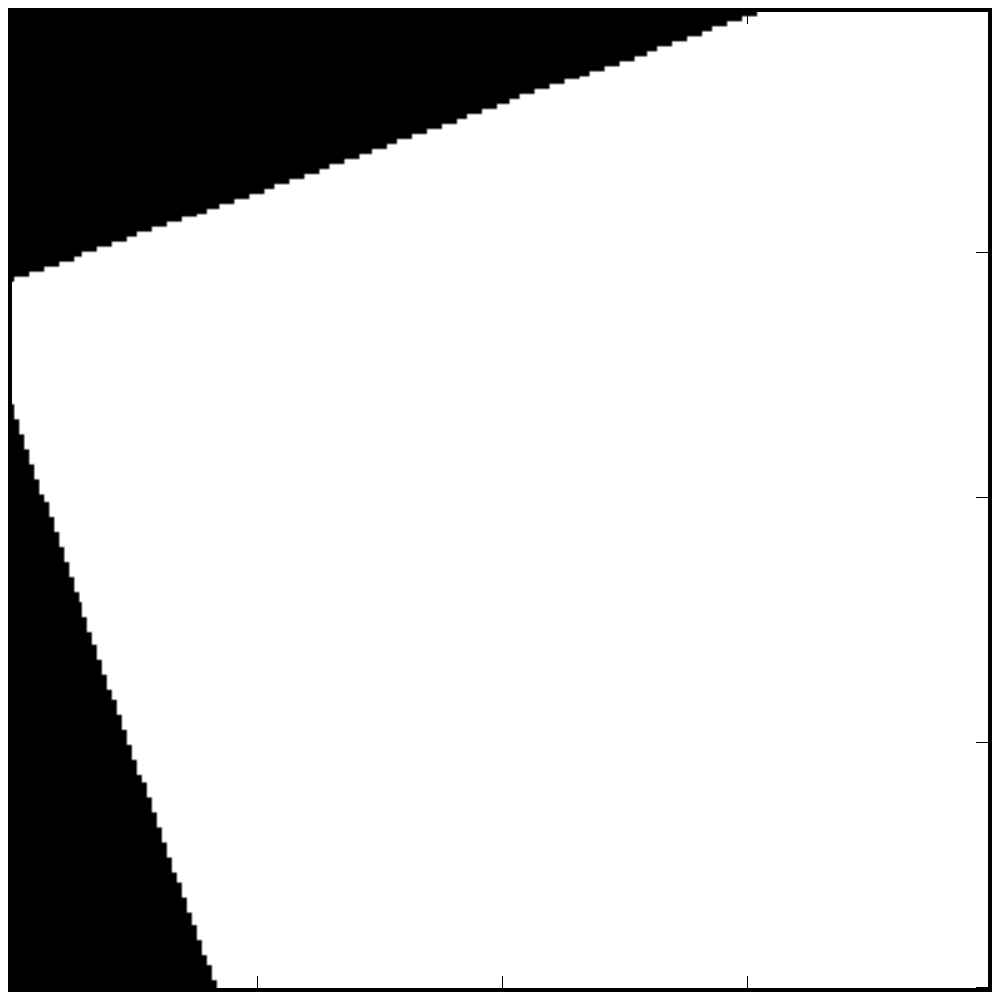} &
\includegraphics[width=0.25\textwidth]{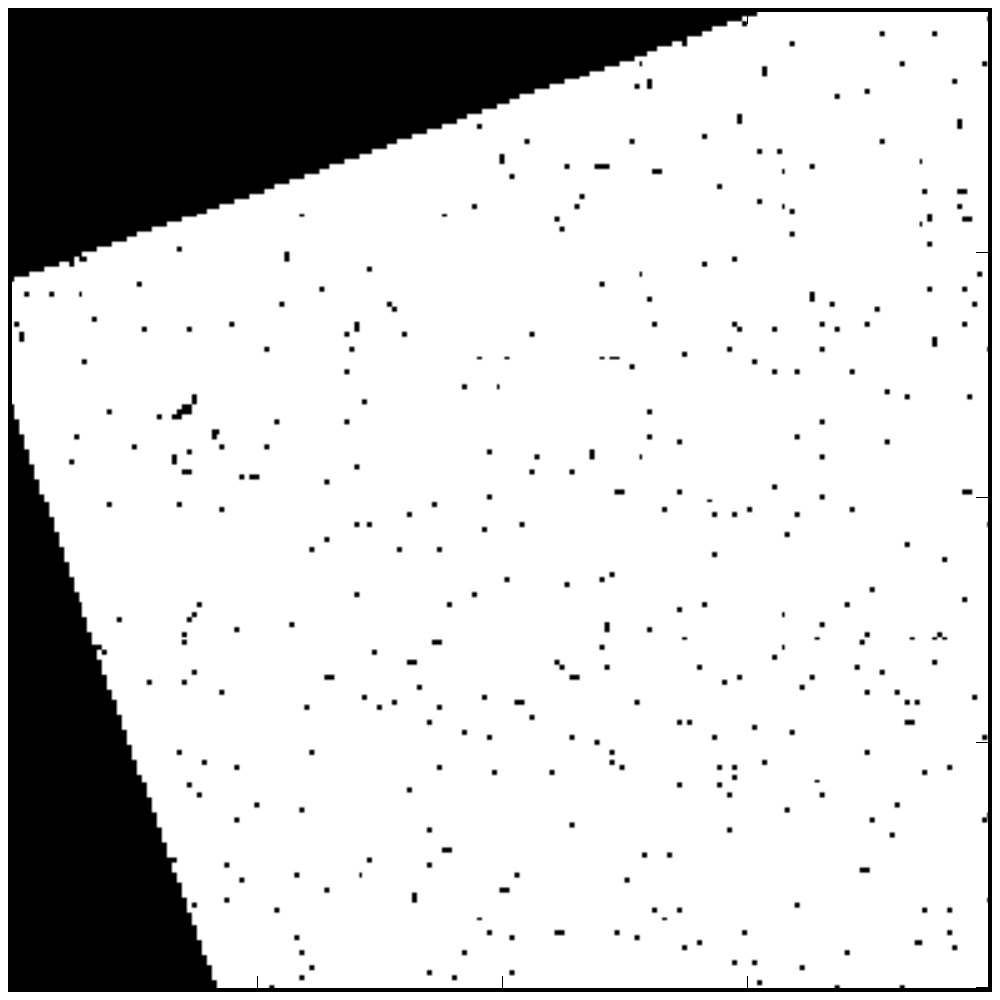} \\
\includegraphics[width=0.25\textwidth]{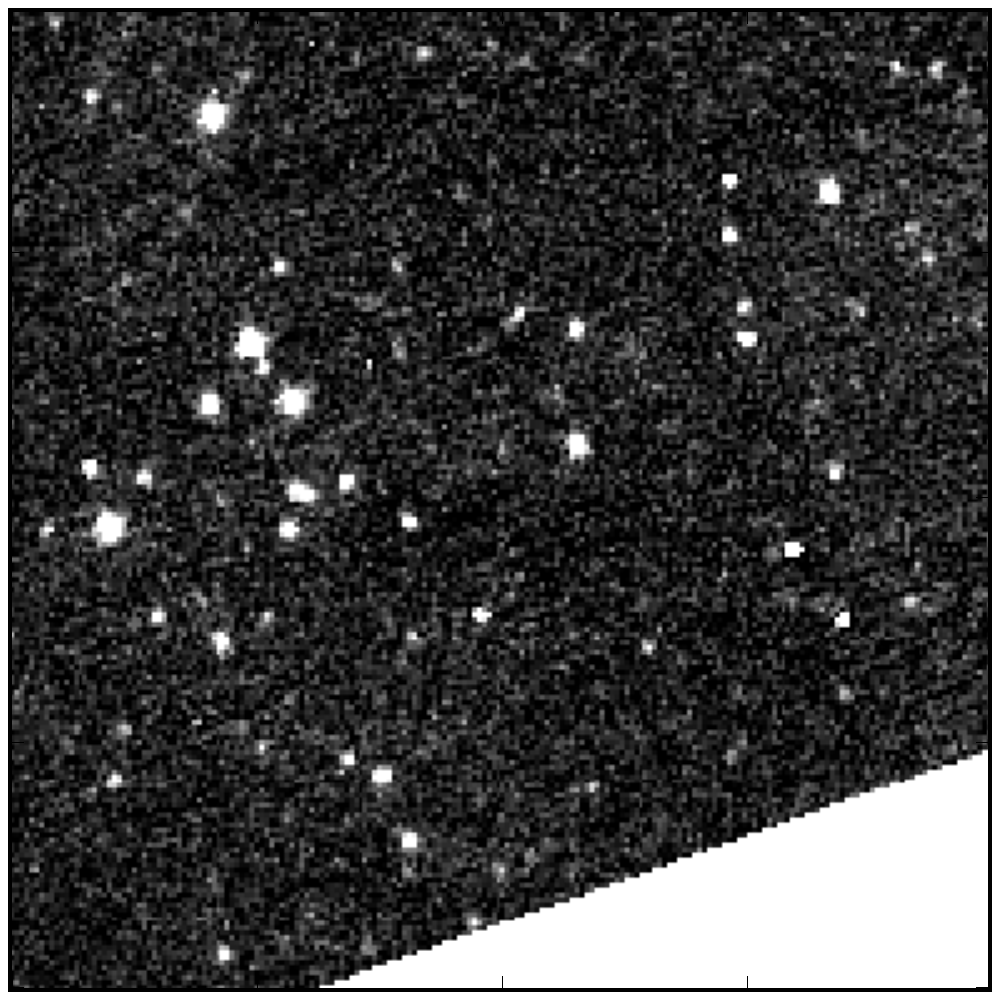} &
\includegraphics[width=0.25\textwidth]{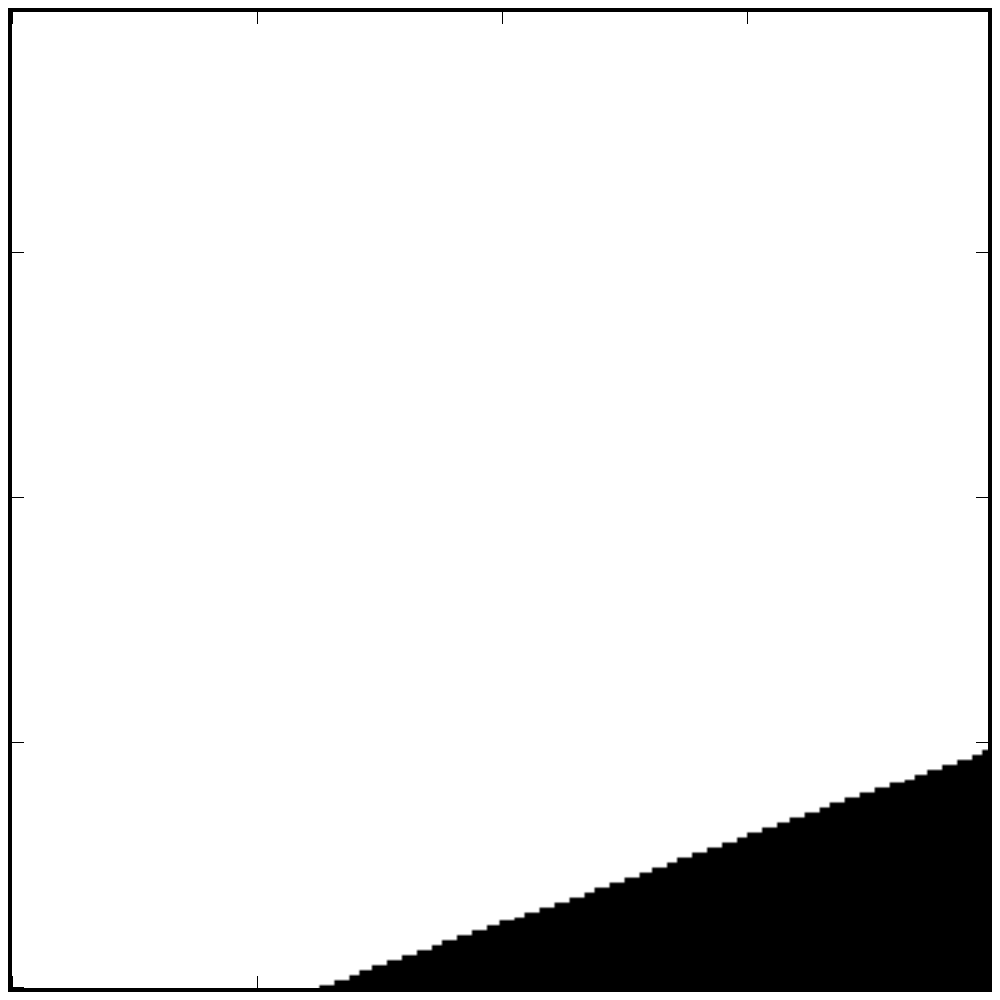} &
\includegraphics[width=0.25\textwidth]{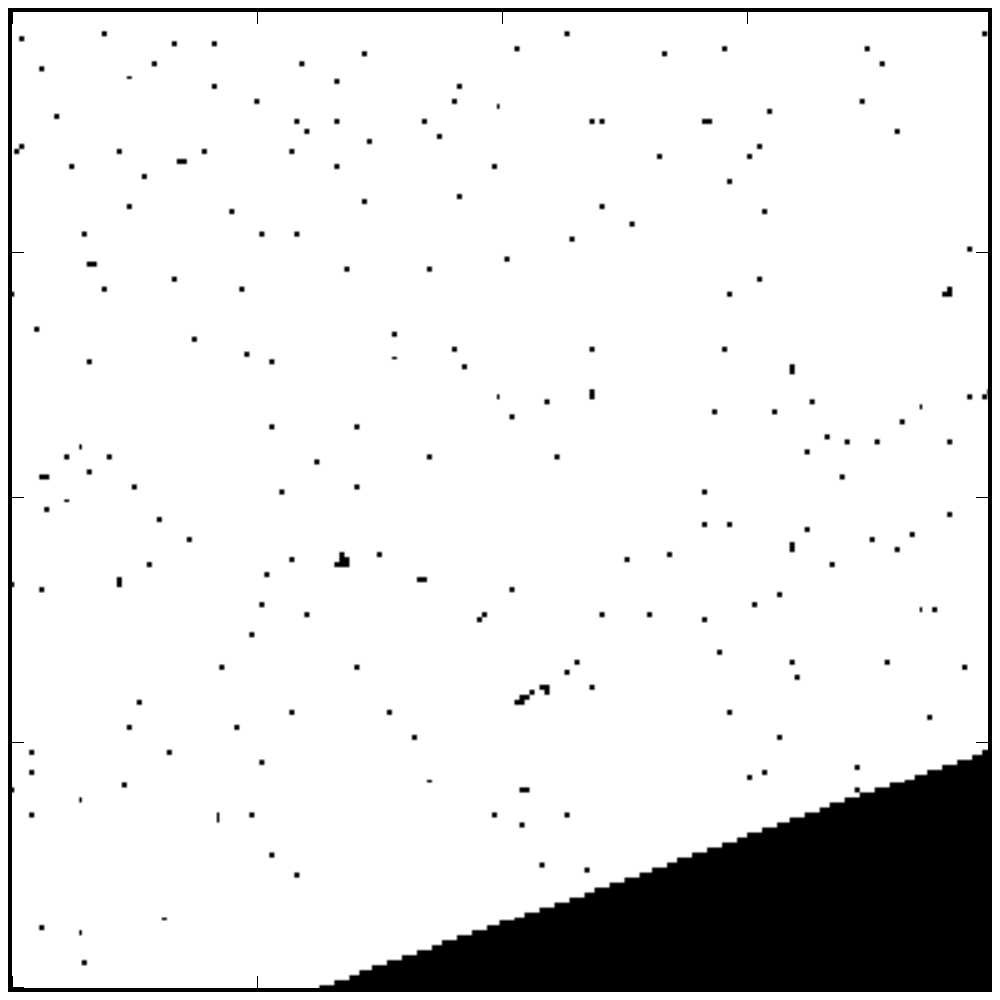}
\end{tabular}
\end{center}
\caption{Image products produced per input frame in the first-round
  coadds, shown for two input frames, for a small
  $200\times200$-pixel cutout in one W1 coadd tile.  \emph{Left:} the
  Lanczos-resampled image.  \emph{Middle:} the binary mask indicating
  which coadd pixels are touched by this input frame.  \emph{Right:}
  the per-pixel masks, nearest-neighbor resampled to the coadd frame.
\label{fig:round1}}
\end{figure}

\begin{figure}
\begin{center}
\begin{tabular}{@{}ccc@{}}
Coadded image ($\coadd$) &
Coadded weights ($\cowt$) &
Per-pixel std.~dev. ($\copp$) \\
\includegraphics[width=0.25\textwidth]{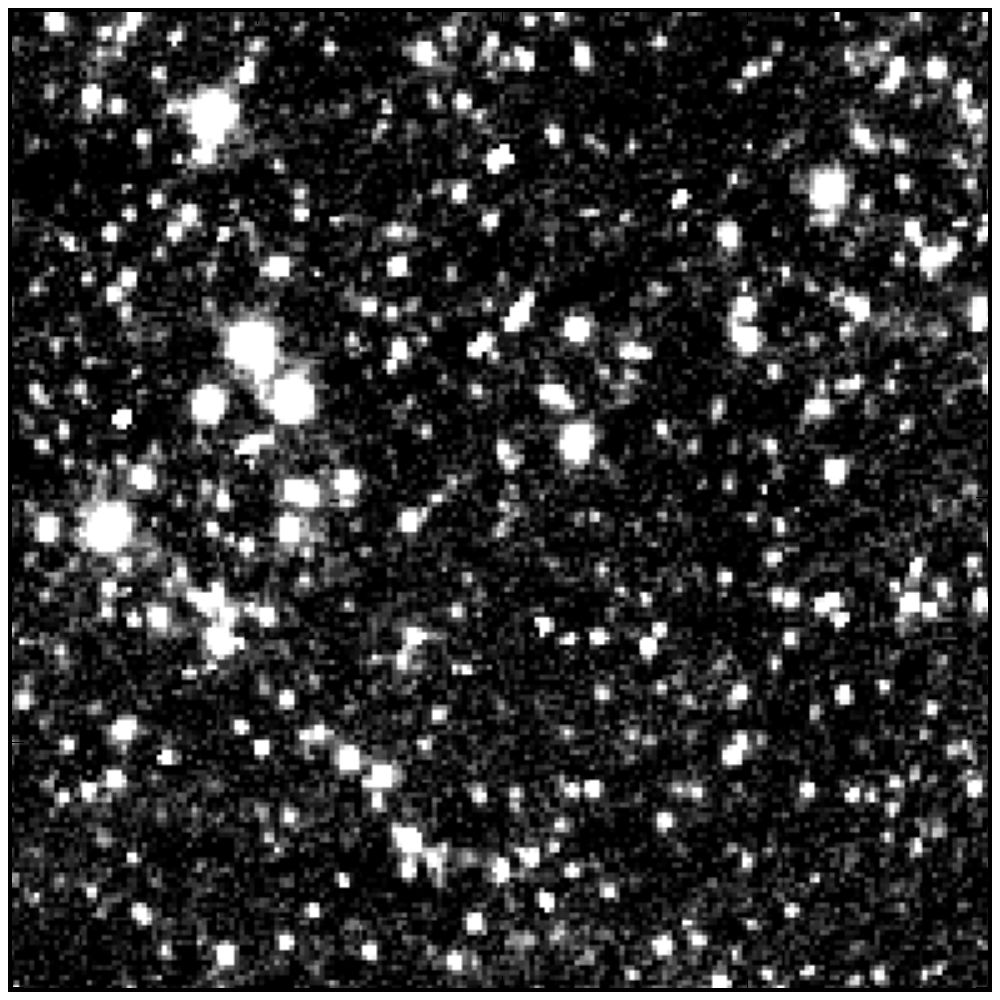} &
\includegraphics[width=0.25\textwidth]{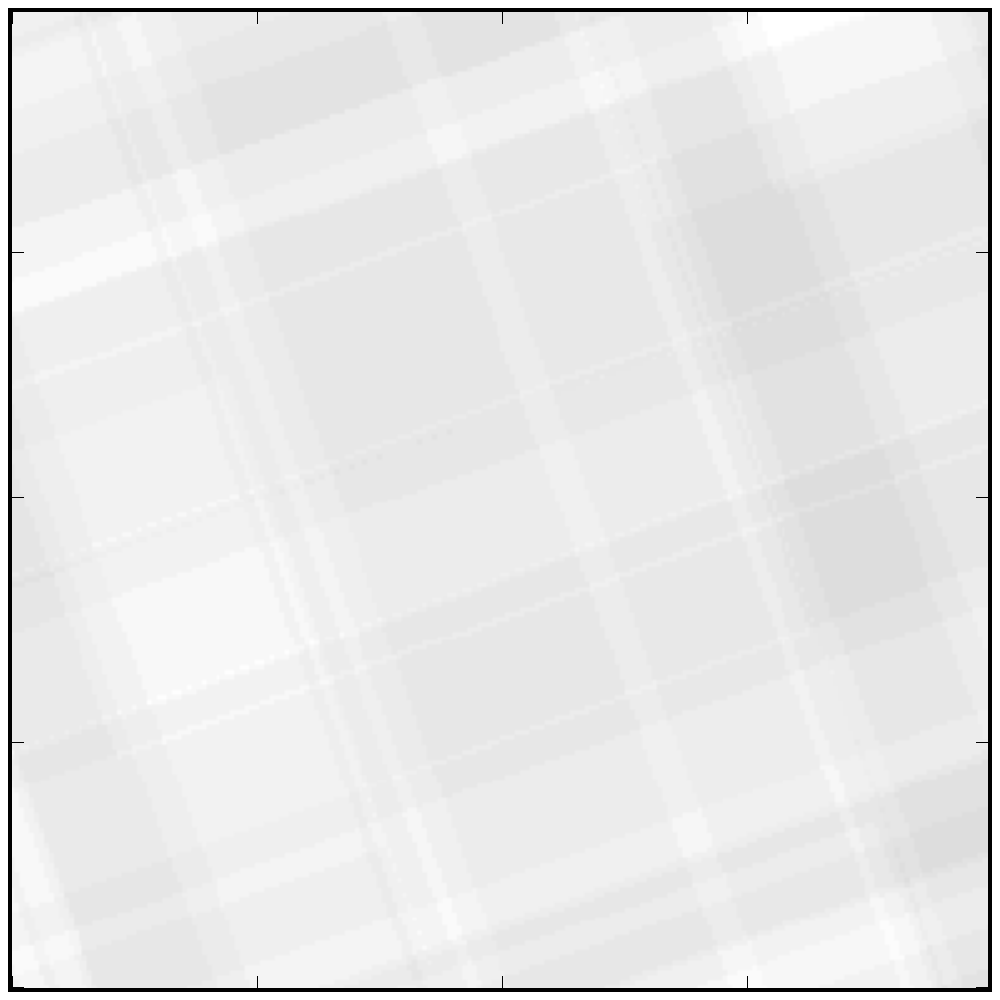} &
\includegraphics[width=0.25\textwidth]{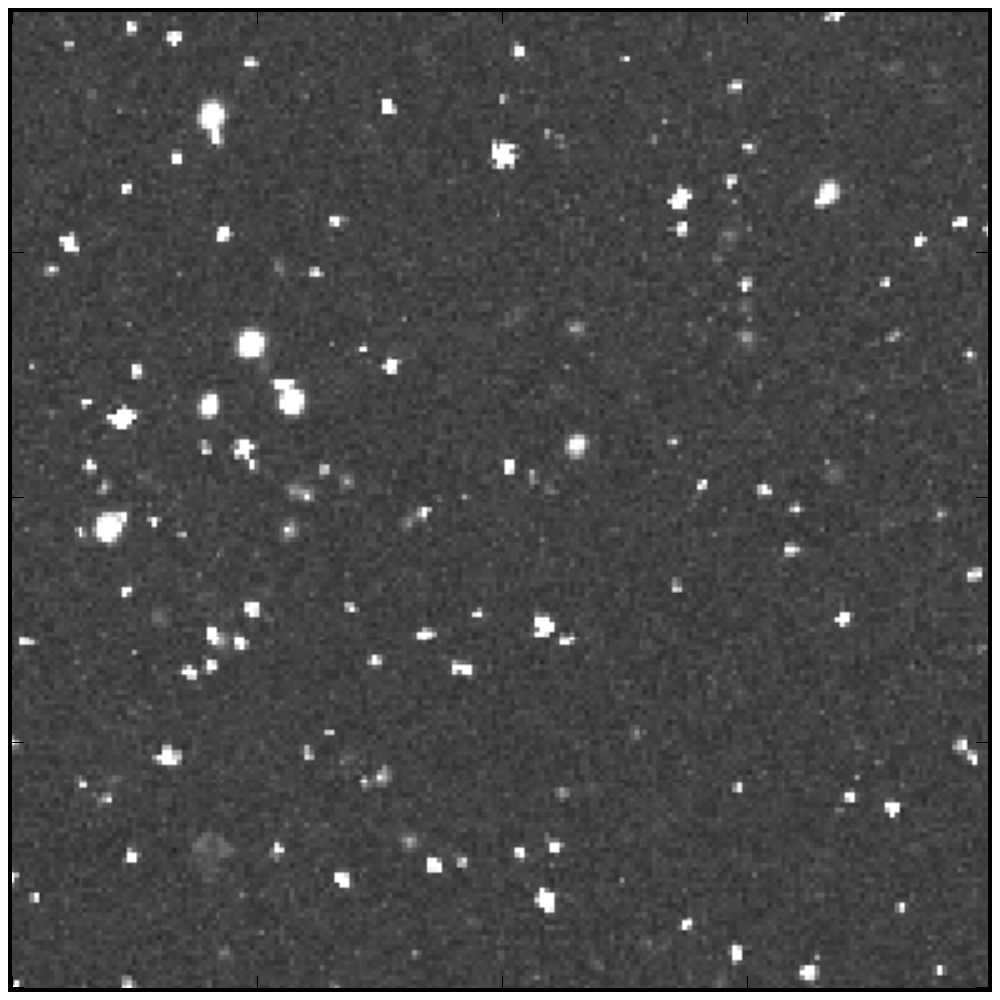}
\end{tabular}
\end{center}
\caption{Coadded image products produced in the first-round, for a
  small $200\times200$-pixel cutout in one W1 coadd tile.  \emph{Left:}
  the coadded intensity image.  \emph{Middle:} the summed weights
  (inverse-variances).  \emph{Right:} the per-pixel standard deviation.
  \label{fig:round1co}}
\end{figure}

\begin{figure}
\begin{center}
\begin{tabular}{@{}ccc@{}}
Coadded image ($\coadd_1$) &
Per-pixel std.~dev. ($\copp_1$) &
\\
\includegraphics[width=0.25\textwidth]{plots1/sequels-024} &
\includegraphics[width=0.25\textwidth]{plots1/sequels-026} &
\\
Resampled image ($\img_i$) & Chi vs coadd ($\ppchi_i$) & Outlier pixels \\
\includegraphics[width=0.25\textwidth]{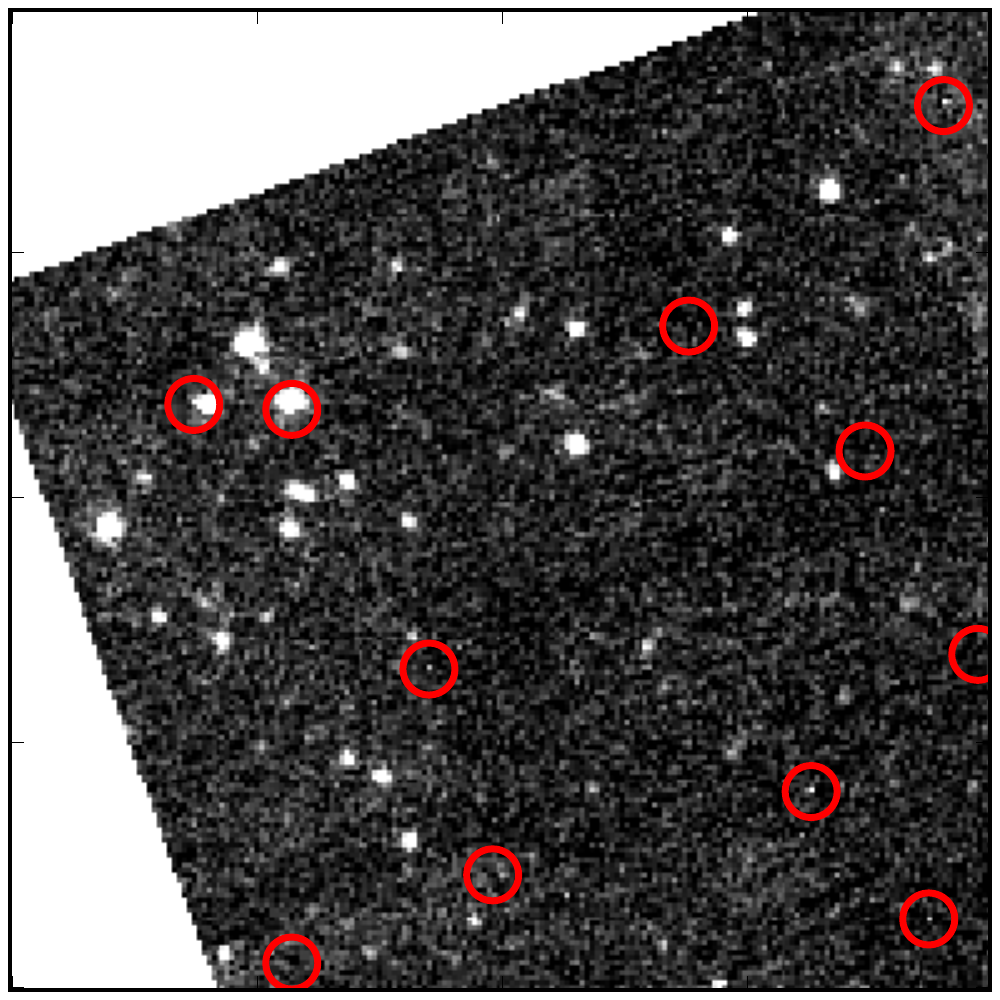} &
\includegraphics[width=0.25\textwidth]{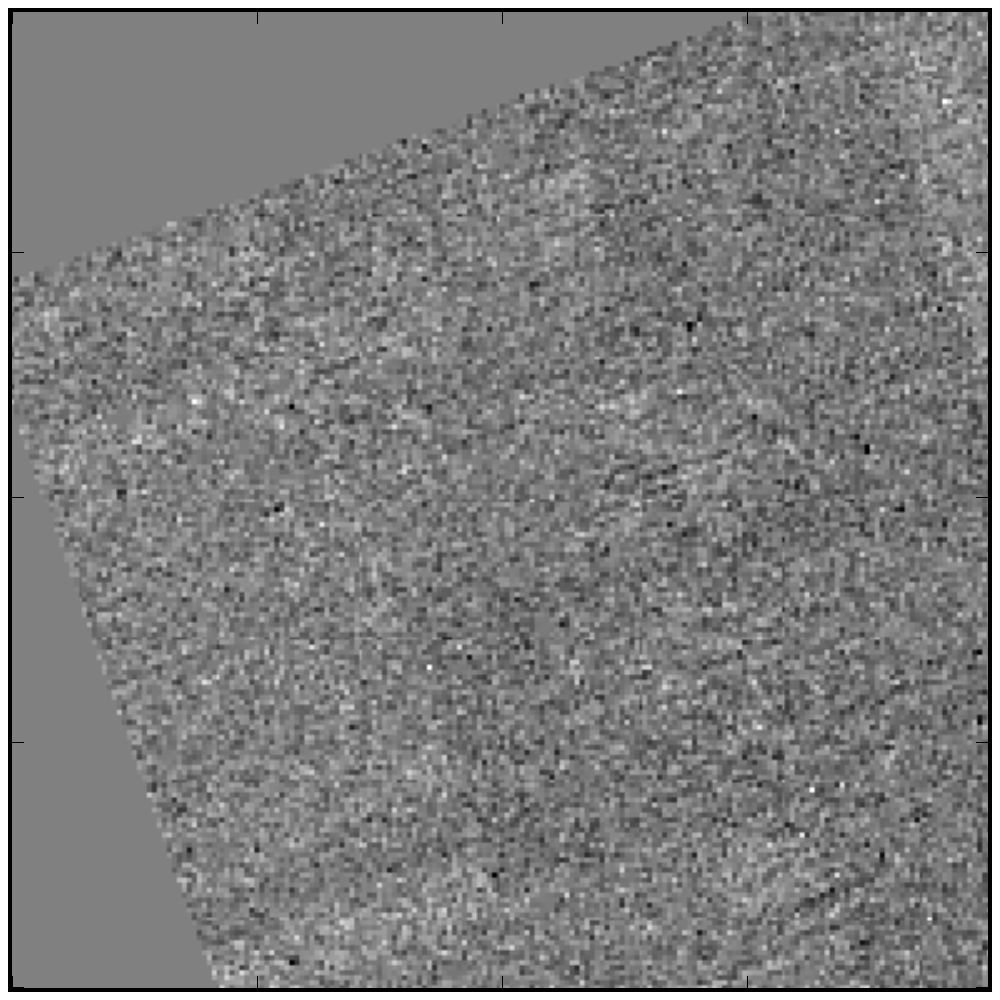} &
\includegraphics[width=0.25\textwidth]{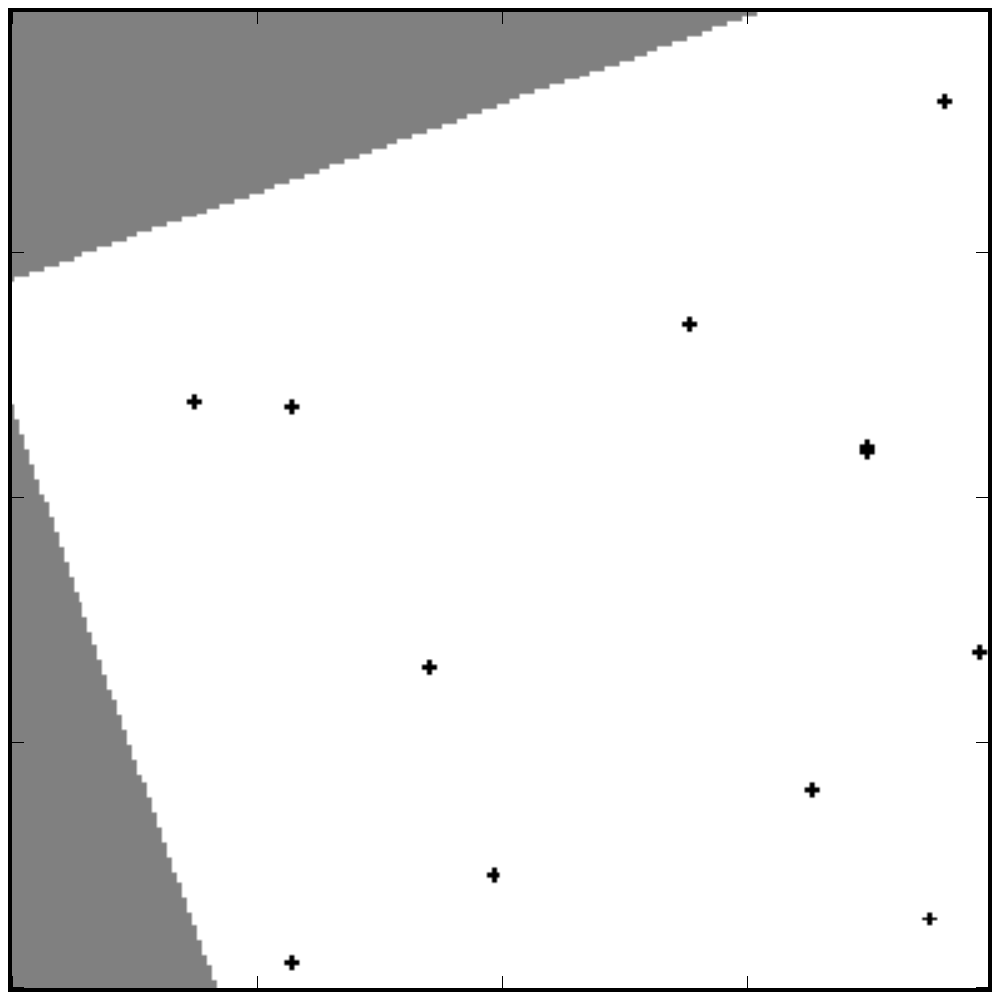} \\
\includegraphics[width=0.25\textwidth]{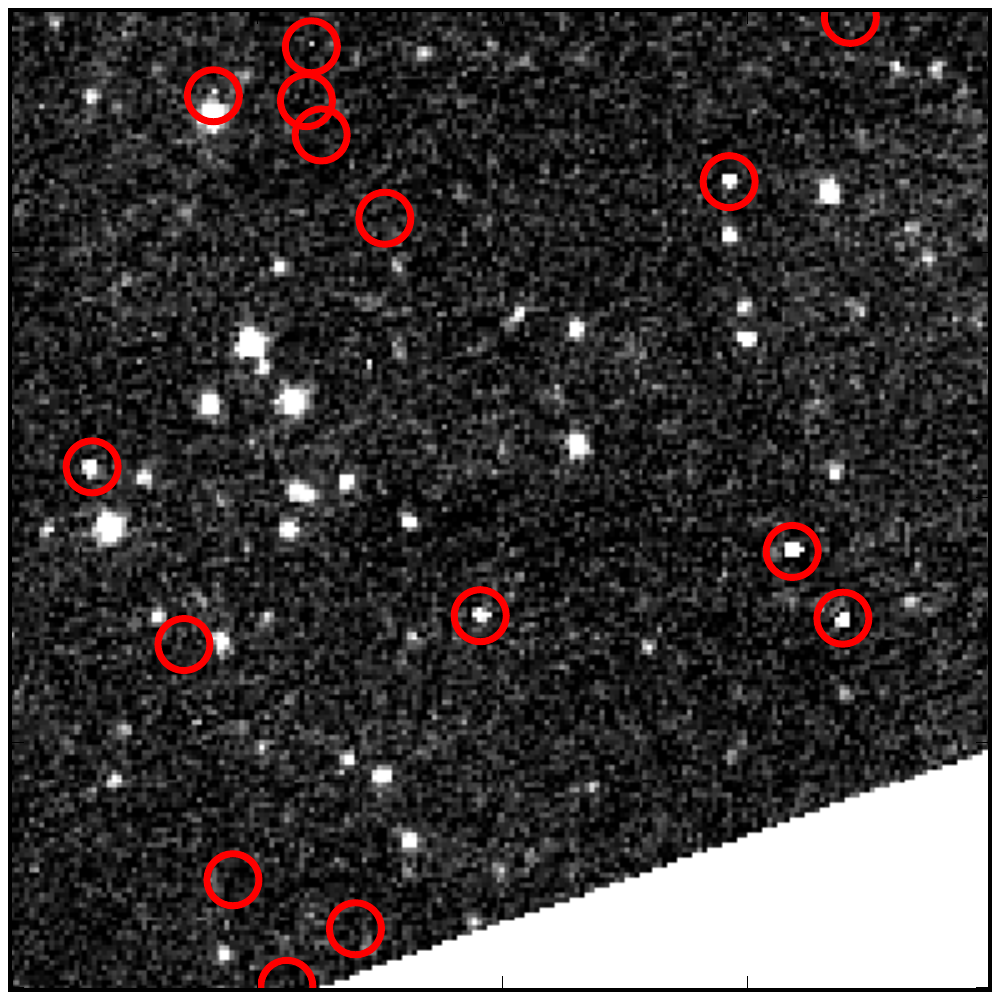} &
\includegraphics[width=0.25\textwidth]{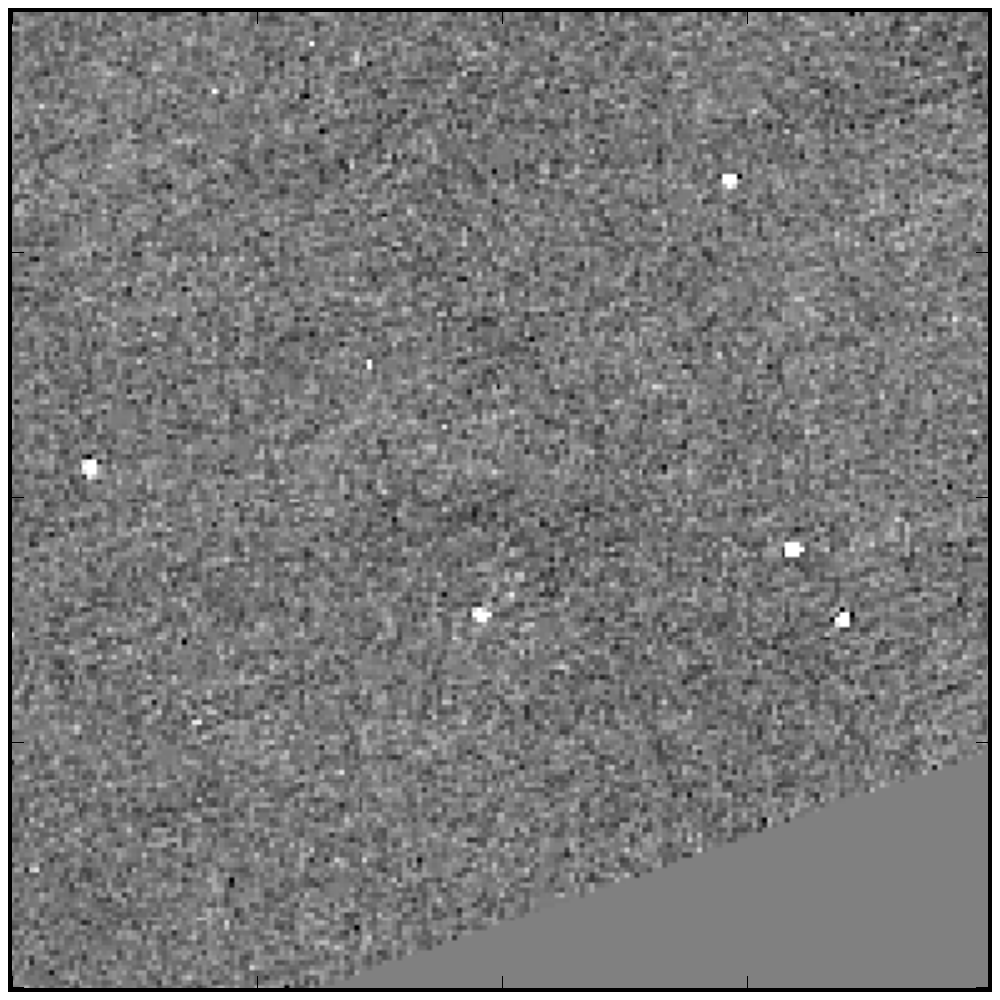} &
\includegraphics[width=0.25\textwidth]{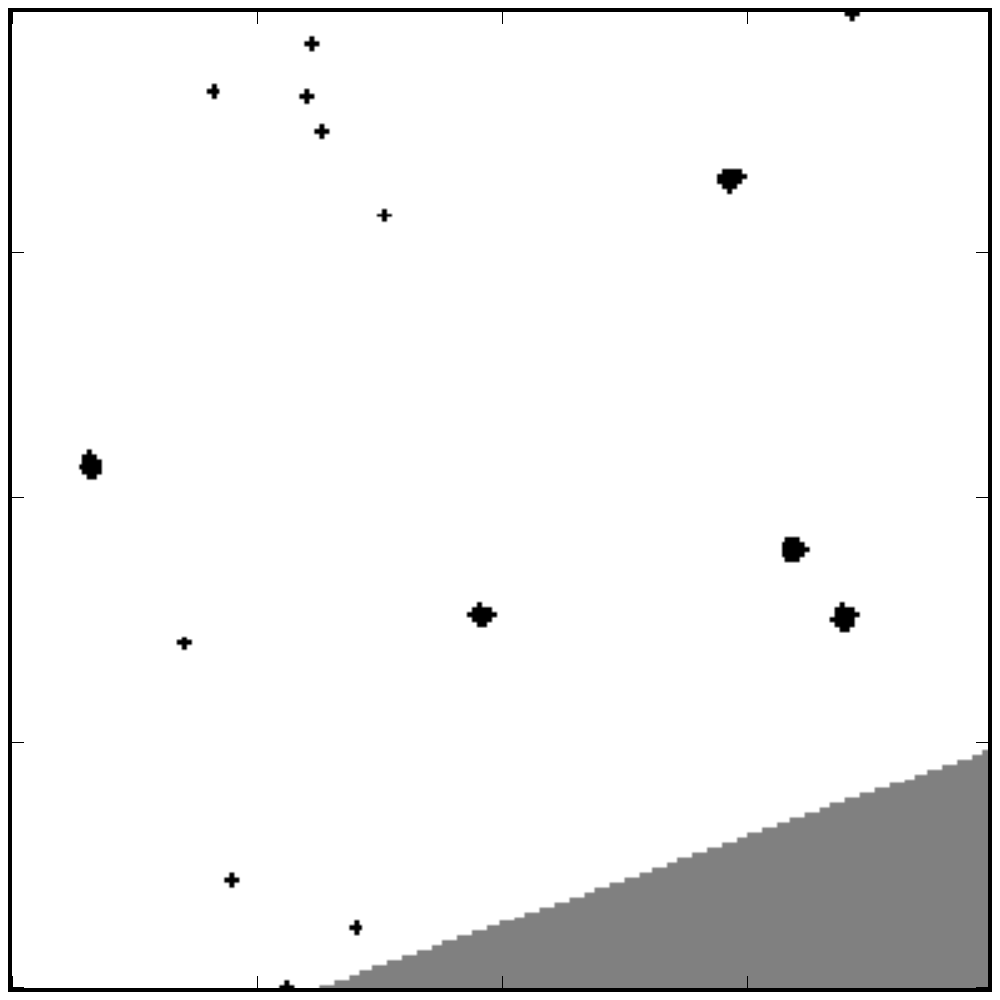}
\end{tabular}
\end{center}
\caption{Outlier pixel detection using per-pixel sample statistics.
  This is the same small W1 region as in \figurename s
  \ref{fig:round1} and \ref{fig:round1co}.  \emph{Top row:}
  First-round coadd and per-pixel standard deviation.  Thanks to
  Poisson statistics, the per-pixel standard deviation is larger where
  sources are found.  This avoids falsely marking too many pixels near
  sources as outliers.  \emph{Middle and Bottom rows, left:} Resampled
  images, with the locations of detected outlier pixels marked.
  Pixels near sources are occasionally falsely mark as outliers.
  \emph{Center:} Chi: resampled image minus coadd mean, divided by
  per-pixel standard deviation (after removing image $i$ from the mean
  and standard deviation estimates).  The large blemishes in the
  bottom row are readily identified.  \emph{Right:} Pixels with
  absolute value of chi greater than 5 are marked as outliers.
\label{fig:round2}}
\end{figure}

I then read each frame in turn, along with its uncertainty image and
masked-pixels image.  I scale each intensity image and its
uncertainty image, using the WISE pipeline zeropoints, to have a
zeropoint of $22.5$.
This is, a source with integrated flux of unity in these images (and
thus the coadds) corresponds to a Vega magnitude of $22.5$.
I estimate an image-wide standard deviation, $\sigma_i$, based on the
median of unmasked pixels in the uncertainty image.  The uncertainty
images (produced by the WISE pipeline) include read noise and
pixelwise Poisson noise from sources, but I want a single image-wide
scalar to avoid biases; the median yields roughly the read noise plus
sky background noise.  I use the inverse-variance as the weight
$\wt_i$ for image $i$; weight $\wt_i = {\sigma_i}^{-2}$.

I find that many of the W3 and W4 images have spatially varying
background levels that appear to be instrumental rather than
astrophysical.  I apply a simple approximate spatial median filter on
the input frames in order to remove some of this background variation.
A full median filter is too expensive so I compute the median in
sub-images of side length $101$ pixels, then interpolate between the
sub-image medians using a piecewise parabolic kernel.  This will of
course also remove astrophysical spatially varying backgrounds, which
leads to severe artifacts in and around extended structures such as
large galaxies and nearby star-forming regions.  This is discussed
below and illustrated in \figurename s \ref{fig:medfilt} and
\ref{fig:badmedfilt}.

In order to resample the images, I \emph{patch} masked pixels, in a
crude manner.  For each masked pixel, I take the average of its four
neighboring pixels, ignoring masked neighbors.  I iterate this
process until no pixels remain to be patched.  The resulting images
have reasonable values in all masked pixels.  As a result, I do not
need to handle masked pixels specially in the interpolation step; if
I did not patch the images, then a single masked pixel would
invalidate an image region the size of the support of the
interpolation kernel (dozens of pixels).  Patching the images allows
me to perform the interpolation with a small approximation error.
I propagate the masks (via nearest-neighbor resampling) to
avoid including pixels that are dominated by patched values in the
coadd.

% http://wise2.ipac.caltech.edu/docs/release/allsky/expsup/sec4_4a.html#maskdef

Next, I estimate and subtract a (scalar) background level from each
frame.  I have taken a somewhat different approach than the WISE
team, based on finding the mode of the pixel value distribution.
Details are given in \secref{sec:impl} below.  For W3 and W4, which
have already had a median-filtered background subtracted, this
additional background subtraction has little effect.

Next, I resample the input frames onto the output coadd frame.  As
mentioned above, the sinc interpolation kernel can be used to perform
this interpolation without loss of information.  The sinc has infinite
extent, so I use the third-order Lanczos kernel as an approximation
(see \secref{sec:impl} below).  For each input frame, I compute a
number of interpolated values: The image itself
(Lanczos-interpolated), $\img_i$, a binary image indicating which
coadd pixels are touched by input pixels (nearest-neighbor
interpolated), $\mask_i$, and a binary image indicating which input
pixels were unmasked, or ``good'' (also nearest-neighbor
interpolated), $\good_i$; see \figref{fig:round1}.  From these, I
accumulate three images:
\begin{align}
  \covar_w  &= \sum_i \img_i^2 \mask_i \wt_i \\
  \coacc_w  &= \sum_i \img_i   \mask_i \wt_i \\
  \cowt     &= \sum_i \mask_i  \wt_i
\end{align}
where $\covar_w$ is the weighted variance image, $\coacc_w$ the
weighted coadd, and $\cowt$ the sum of (masked) weights.  Recall that
$\img_i$ and $\mask_i$ are images, while $\wt_i$ is a scalar (the
image-wide inverse variance described above).  The first-round coadd
$\coadd_1$ and the per-pixel sample standard deviation $\copp_1$, are:
\begin{align}
  \coadd_1  &= \frac{\coacc_w}{\cowt} &
  \copp_1   &= \sqrt{\frac{\covar_w}{\cowt} - \coadd_1^2} \quad;
\end{align}
see \figref{fig:round1co}.

Using the per-pixel sample variance computed in the first-round coadd,
I detect pixels that are significant outliers.  In order to avoid
outliers themselves biasing the estimated sample mean and variance, I
have to remove each frame's contribution from the sample mean (coadd)
and sample variance,
\begin{align}
\coadd_i  &= \frac{\coacc_w - \img_i   \mask_i \wt_i}{\cowt - \mask_i \wt_i}
&
\copp_i^2 &= \frac{\covar_w - \img_i^2 \mask_i \wt_i}{\cowt - \mask_i \wt_i} - \coadd_i^2%
 \quad .
\end{align}
Furthermore, since the sample variance estimate given a small number
of samples can be noisy, I apply a prior on the variance,
$\sigprior_i^2$, to get a regularized variance estimate $\hat{\copp}_i^2$:
\begin{align}
\sigprior_i^2 &= \sigma_i^2 + (\fluxsys \coadd_i)^2 \\
\hat{\copp}_i &= \sqrt{%
  \frac{\copp_i^2 (W - \mask_i \wt_i) + \sigprior_i^2 \, \nprior \wt_i}%
       {W - \mask_i \wt_i + \nprior \wt_i}%
       }
\end{align}
where $\fluxsys$ is a multiplicative flux error (set to $3\%$; see
\secref{sec:results}), and $\nprior$ is the effective weight of the
prior (set to $5$); this is similar to a Gamma prior.
Using these estimates of the mean and standard deviation, I compute
the per-pixel chi value,
\begin{align}
\ppchi_i &= \frac{\img_i - \coadd_i}{\hat{\copp}_i} \quad ,
\end{align}
and mask any pixel $i$ with $\abs{\ppchi_i} > 5$ as an outlier.  See
\figref{fig:round2}.  This readily removes artifacts such as cosmic
rays and satellite trails, and will also remove transient or
fast-moving sources.  I mask the 4-connected neighbors of masked
pixels as well, resample these masks back to the original image frame
via nearest-neighbor interpolation, and record them for future use.  I
discard any frame that has more than 1\% of its pixels masked.  I then
patch the frame's resampled image and accumulate it into the
second-round coadds.

During the second round, I patch outlier pixels, and add the outliers
to the pixel masks:
\begin{align}
\simg_i  &= \patch{\img_i}{\abs{\chi_i} > 5}
&
\sgood_i &= \good_i \binaryand (\abs{\chi_i} \le 5)
\end{align}
(where $\binaryand$ denotes binary AND) and then proceed to accumulate
``unmasked'' and ``masked'' values:
\begin{align}
\covar_u &= \sum_i {\simg}_i^2 \mask_i \wt_i 
&
\covar_m &= \sum_i {\simg}_i^2 \sgood_i \wt_i
\\
\coacc_u &= \sum_i \simg_i     \mask_i \wt_i 
&
\coacc_m &= \sum_i \simg_i     \sgood_i \wt_i
\\
\cowt_u  &= \sum_i             \mask_i \wt_i 
&
\cowt_m  &= \sum_i             \sgood_i \wt_i
\\
\con_u   &= \sum_i             \mask_i       
&
\con_m   &= \sum_i             \sgood_i
\end{align}
and at the end, the coadd products are:
\begin{align}
\coadd_u  &= \frac{\coacc_u}{\cowt_u}
&
\coadd_m &= \frac{\coacc_m}{\cowt_m}
\\
\copp_u   &= \frac{\sqrt{\frac{\covar_u}{\cowt_u} - \coadd_u^2}}{\sqrt{\con_u - 1}}
&
\copp_m   &= \frac{\sqrt{\frac{\covar_m}{\cowt_m} - \coadd_m^2}}{\sqrt{\con_m - 1}}
\end{align}
where the difference between the ``masked'' $\coadd_m$ and
``unmasked'' $\coadd_u$ is that the masked versions ignore masked
pixels in the input images (including pixels flagged as outliers),
while the unmasked versions use the ``patched'' pixel values for
masked pixels rather than ignoring them altogether.  The per-pixel
sample standard deviations $\copp_m$ and $\copp_u$ are scaled so that
they are approximate error estimates for the coadd intensity pixels.

Finally, I re-estimate and subtract a scalar background level from
the final coadds.  This might seem to be unnecessary, since I
subtracted the background from the input frames going into the coadd.
But the coadds have considerably higher signal-to-noise; a pixel with
one sigma of source flux in the individual frames will appear to be a
blank pixel affected only by the background, but in the coadd it will
have several sigma of source flux, and will no longer be considered
part of the ``blank'' sky that defines the background level.  In this
way, the coadd's pixel distribution preferentially shifts pixels with
positive flux out of the ``blank sky'' peak into the ``source'' tail,
meaning that the peak of the remaining ``blank sky'' pixel
distribution is shifted negative.

The output products include, for each coadd tile: \emph{(a)} co-added
intensity images that exclude masked outliers and masked input pixels,
$\coadd_m$, as well as ones that replace these masked pixels by patching,
$\coadd_u$; \emph{(b)} pixelwise inverse-variance maps $\cowt_m$ and
$\cowt_u$; \emph{(c)} per-pixel sample standard deviation maps
$\copp_m$ and $\copp_u$; \emph{(d)} number-of-frames coverage maps
$\con_m$ and $\con_u$; \emph{(e)} FITS tables summarizing the frames
considered for inclusion in the coadd; and \emph{(f)} mask images for
the input-frame pixels flagged as outliers.

\subsection{Implementation Details}
\label{sec:impl}

I estimate a scalar background level in each input frame as follows.
I attempt to find the mode of the image pixel distribution by
histogramming image pixels in a region around the peak and fitting a
parabola to the log-counts in each histogram bin; this assumes a
Gaussian distribution of pixel values around the mode.  I first
produce a coarsely-binned pixel histogram, find the largest bin, then
find the range of bins containing more than 50\% as many pixels (on
the low side) and 80\% as many pixels (on the high side).  I
histogram this region more finely and fit a parabola to the
log-counts.  This region has many counts in each histogram bin and is
not too contaminated by pixels with significant source flux.  The WISE
pipeline's dynamic calibration (\emph{dynacal}) stage is
known\footnote{Eg,
  \niceurl{http://wise2.ipac.caltech.edu/docs/release/allsky/expsup/sec8\_3b.html\#dyna}}
to introduce an artifact in which a large fraction of pixel values are
set to the median value; to alleviate this problem I add Gaussian
noise with variance equal the per-pixel variance before histogramming
the pixel values.

Resampling the input frames onto the output coadd frame requires a
large number of astrometric World Coordinate System (WCS)
transformations.  Since these can be expensive, I use a spline
approximation: I evaluate the WCS on a $25\times25$-pixel grid,
mapping coadd pixel coordinates to input coordinates, and use a cubic
spline to interpolate this mapping.  A similar approach is used in
SWarp \citep{swarp}.

I use third-order Lanczos interpolation to resample the input frames.
This requires a large number of evaluations of the Lanczos kernel (a
windowed sinc function), which can be quite expensive.  I approximate
this using a precomputed look-up table with 2048 bins per unit
interval.  This incurs negligible approximation error in practice.

\section{Results}
\label{sec:results}

\begin{figure}
\begin{center}
\begin{tabular}{@{}ccc@{}}
WISE & unWISE (unmasked) & unWISE (masked) \\
\includegraphics[height=0.25\textwidth]{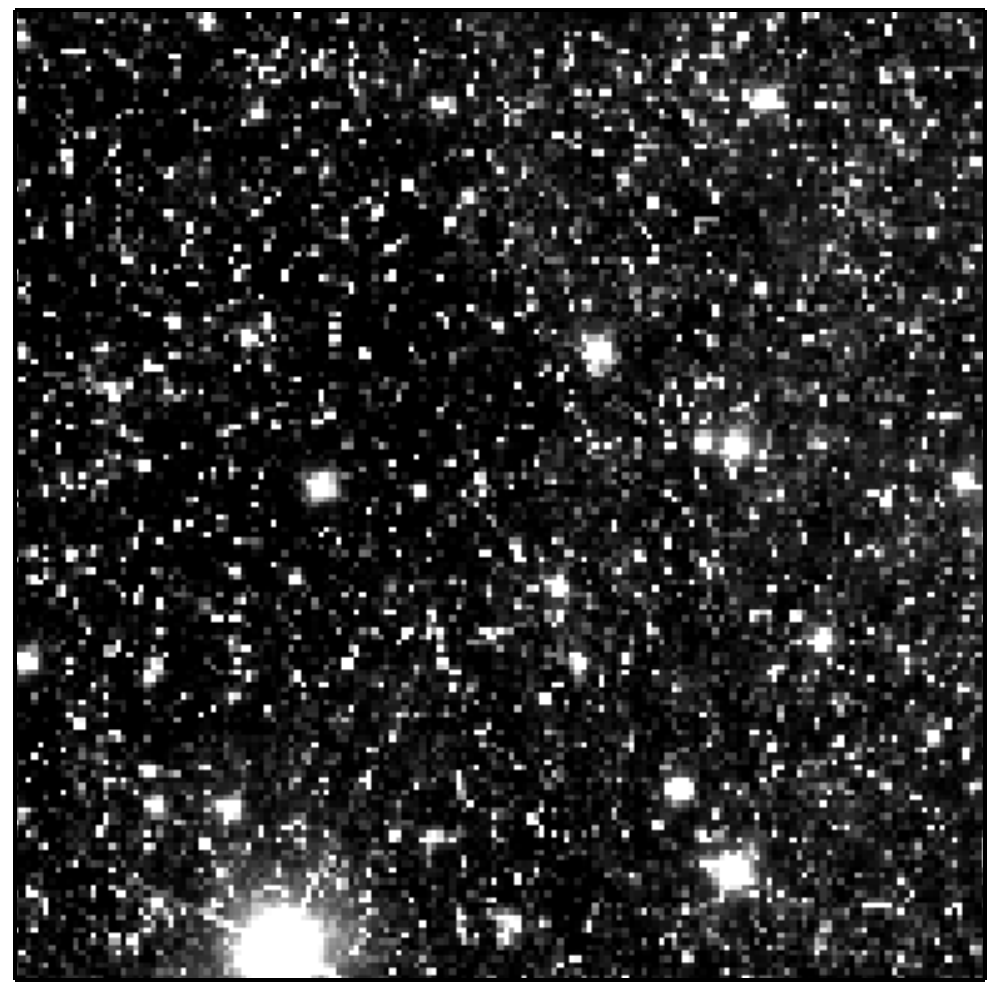} &
\includegraphics[height=0.25\textwidth]{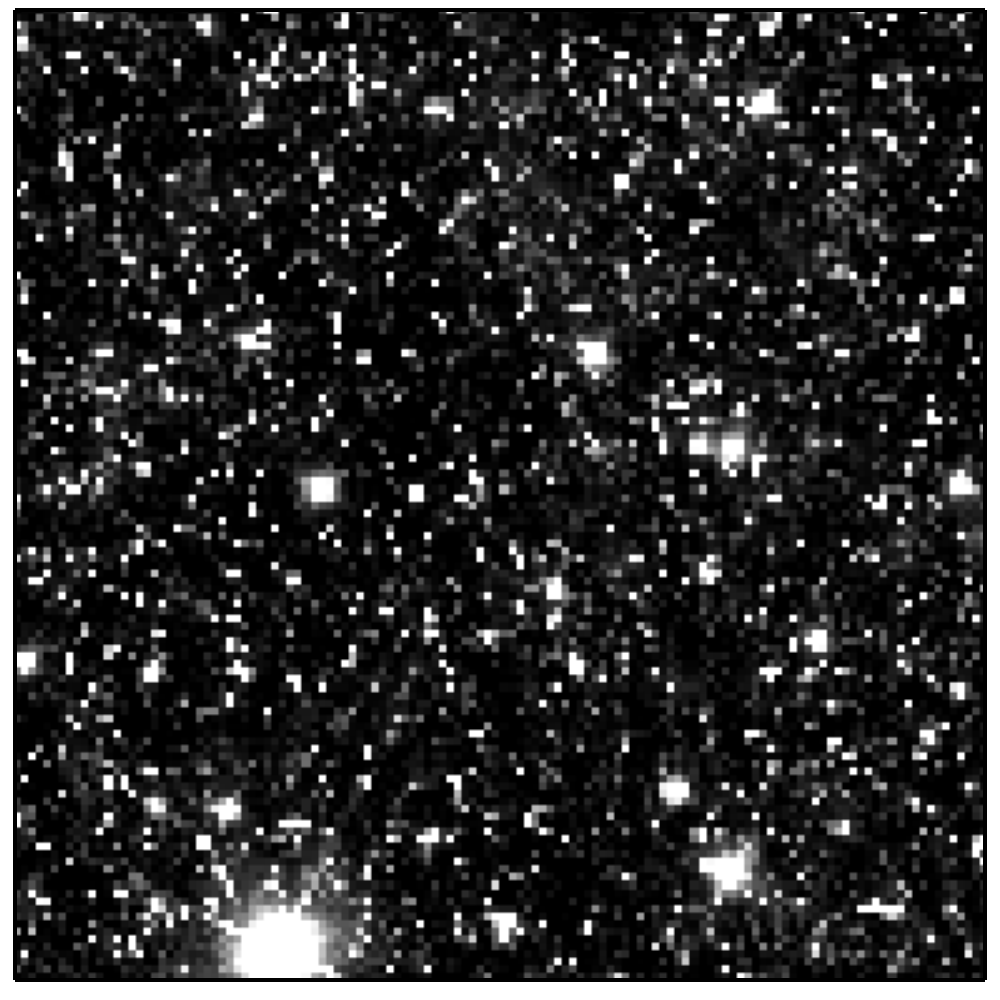} &
\includegraphics[height=0.25\textwidth]{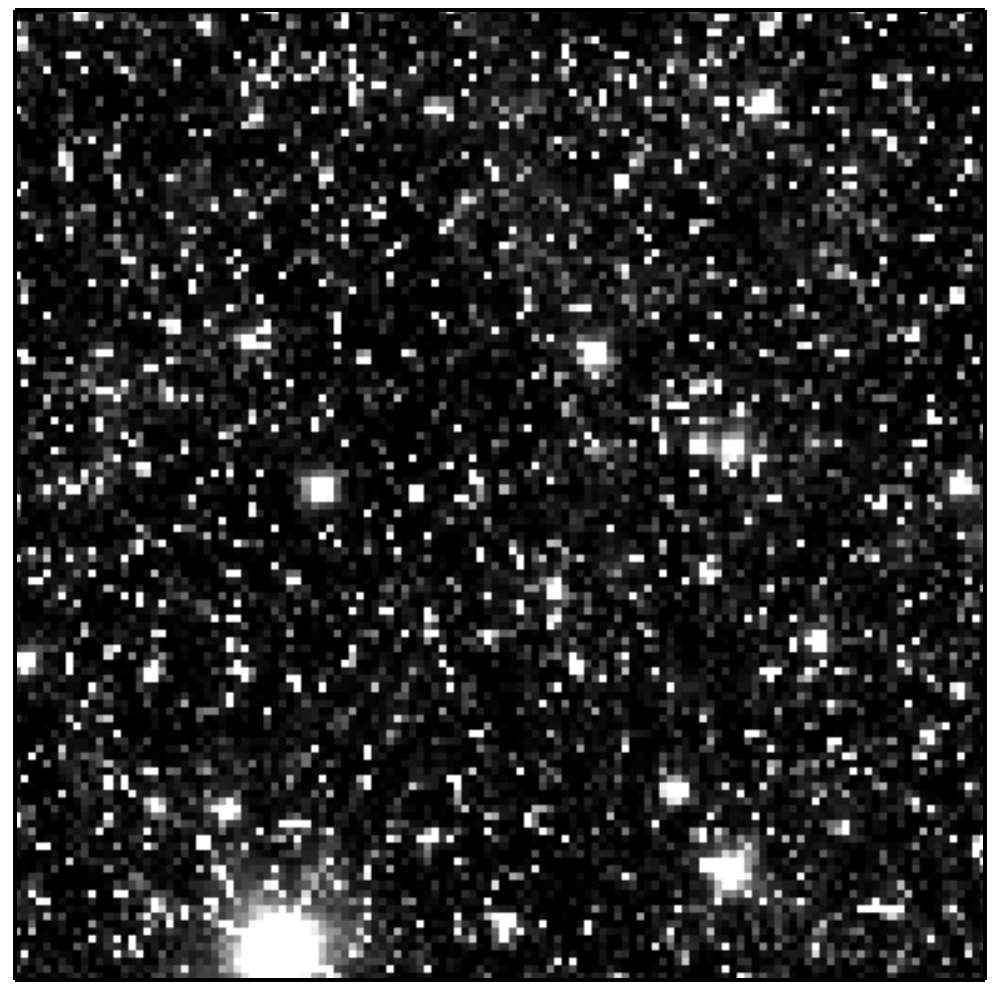} \\
\includegraphics[height=0.25\textwidth]{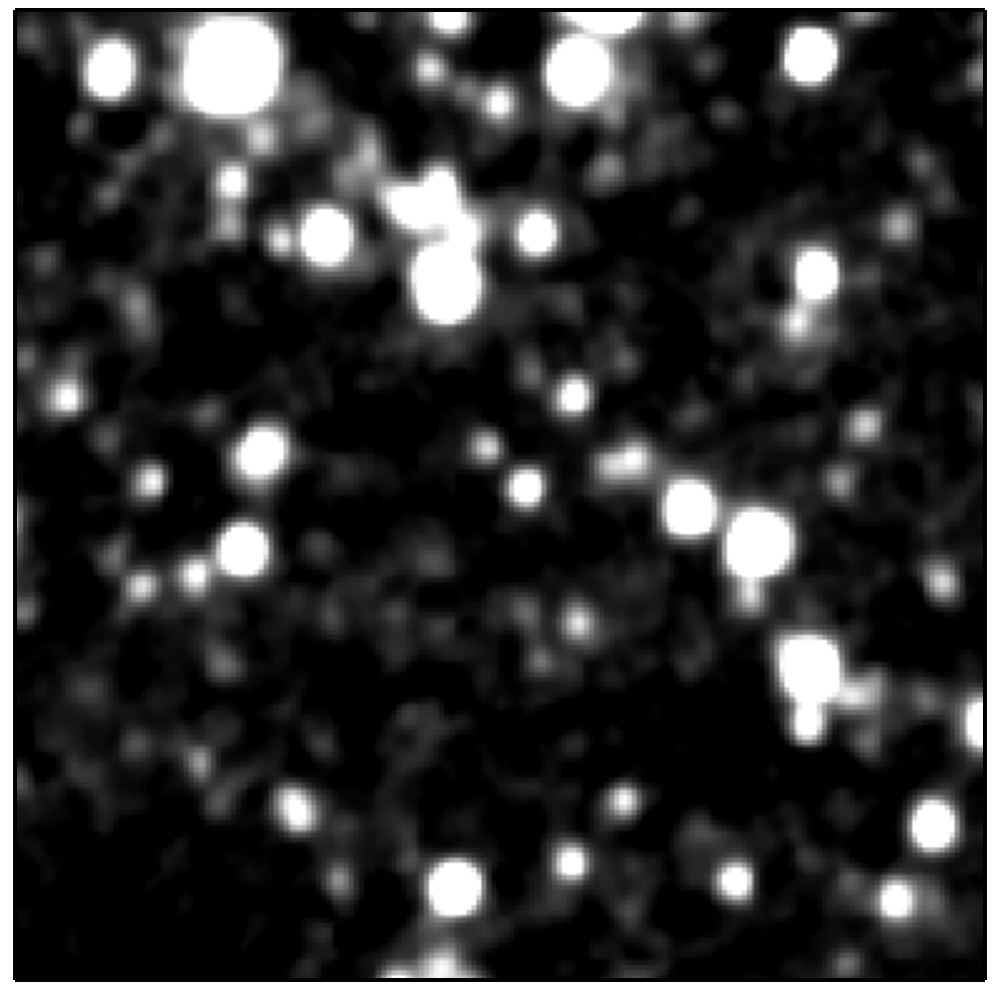} &
\includegraphics[height=0.25\textwidth]{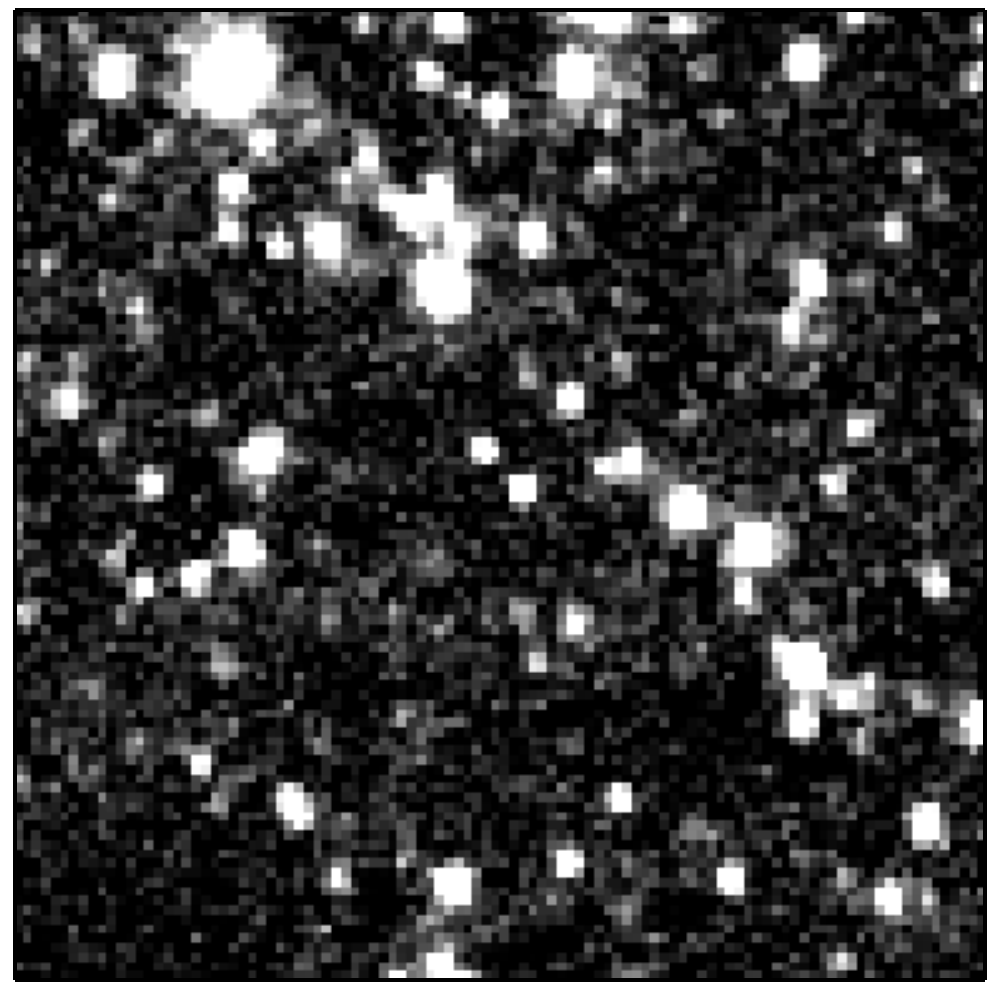} &
\includegraphics[height=0.25\textwidth]{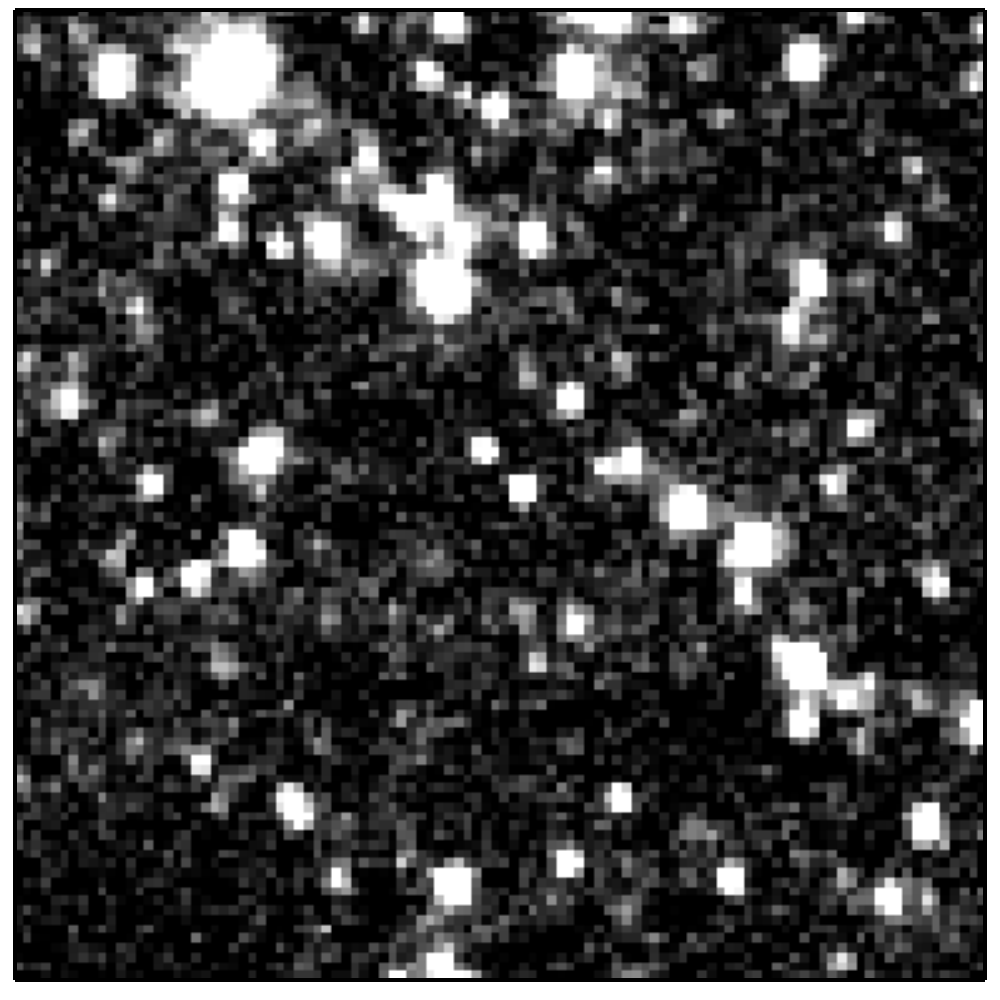} \\
\includegraphics[height=0.25\textwidth]{\bwfig{plots2/co-06}} &
\includegraphics[height=0.25\textwidth]{\bwfig{plots2/co-07}} &
\makebox[0.25\textwidth][l]{\includegraphics[height=0.25\textwidth]{\bwfig{plots2/co-08}}} \\
\end{tabular}
\end{center}
\caption{One example image: coadd tile 1384p454, band W1.  \emph{Top
    row:} Intensity images.  Overall, my coadds and the AllWISE
  Release coadds are very similar.  \emph{Second row:} Zoom-in of
  intensity images to middle 5\% of image.  Here it is clear that the
  AllWISE Release coadds have been artificially blurred, while mine
  maintain full resolution. \emph{Bottom row:} Number of images
  included in coadd.  
\label{fig:res1}}
\end{figure}

An example coadd image is shown in \figref{fig:res1}.

\begin{figure}
\begin{center}
\includegraphics[width=0.8\textwidth]{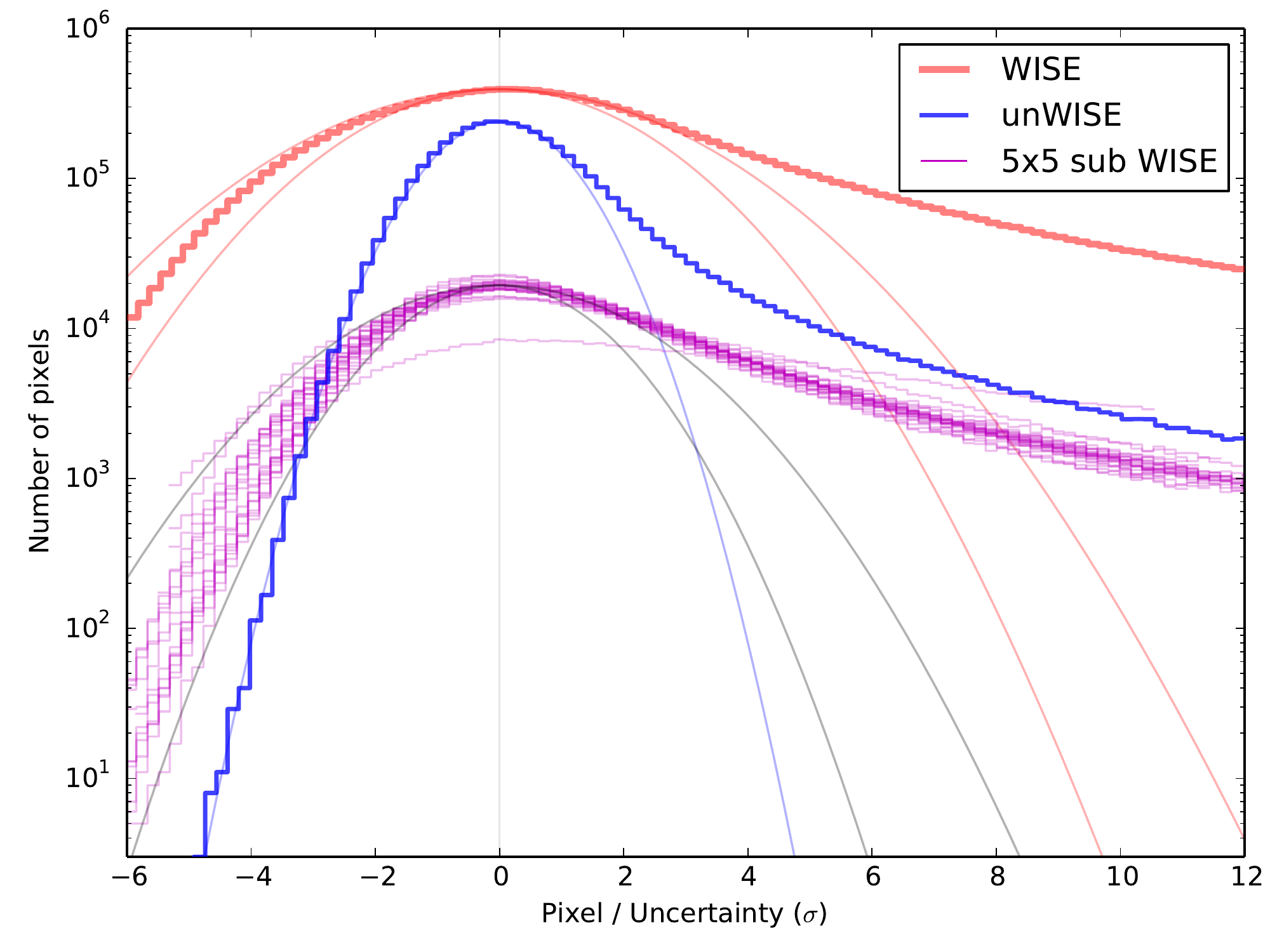}
\end{center}
\caption{Coadd pixel distributions for example tile 1384p454, band W1.
  Each curve shows the coadd intensity values divided by the per-pixel
  uncertainty values.  The broad histogram labeled ``WISE'' shows the
  AllWISE Release \emph{intensity} over \emph{uncertainty} maps, after
  estimating and removing a scalar background using the method
  described above.
  In the absence of sources and varying backgrounds, this should
  approximate a Gaussian with unit standard deviation.  Sources add
  the positive tails, and a varying background would broaden the
  distribution.
  The AllWISE Release pixel distribution is considerably broader than
  expected (the uncertainty maps do not capture the pixel variations)
  by more than a factor of two (the top thin guidelines are
  Gaussians with standard deviations 2 and 2.5).  Splitting the image into
  $5 \times 5$ subimages, the subimage distributions show a slightly
  narrower distribution, though still broader than expected (the
  guidelines show Gaussians with standard deviations $\sqrt{2}$
  and 2); the $20\times20$ subimages (not shown here) are similar.
  This suggests that if background variations are responsible for this
  effect, the relevant scale is smaller than $\sim200$ pixels.
  This effect was less pronounced in the All-Sky Release Atlas Images.
  The curve labeled ``unWISE'' shows the unWISE coadd \emph{image} times square
  root of \emph{inverse-variance} maps, with no scaling or
  adjustments.  This distribution is very close to unit Gaussian (plus
  source tail).  The WISE curves are shifted higher because the WISE
  coadds have a smaller pixel scale and hence more pixels.
  \label{fig:pixdist}}
\end{figure}

\begin{figure}
\begin{center}
\begin{tabular}{@{}c@{\hspace{0.05\textwidth}}c@{}}
WISE & unWISE \\
% I like the B/w versions better anyway
%\includegraphics[width=0.4\textwidth]{\bwfig{plots2/co-12}} &
%\includegraphics[width=0.4\textwidth]{\bwfig{plots2/co-13}}
\includegraphics[width=0.4\textwidth]{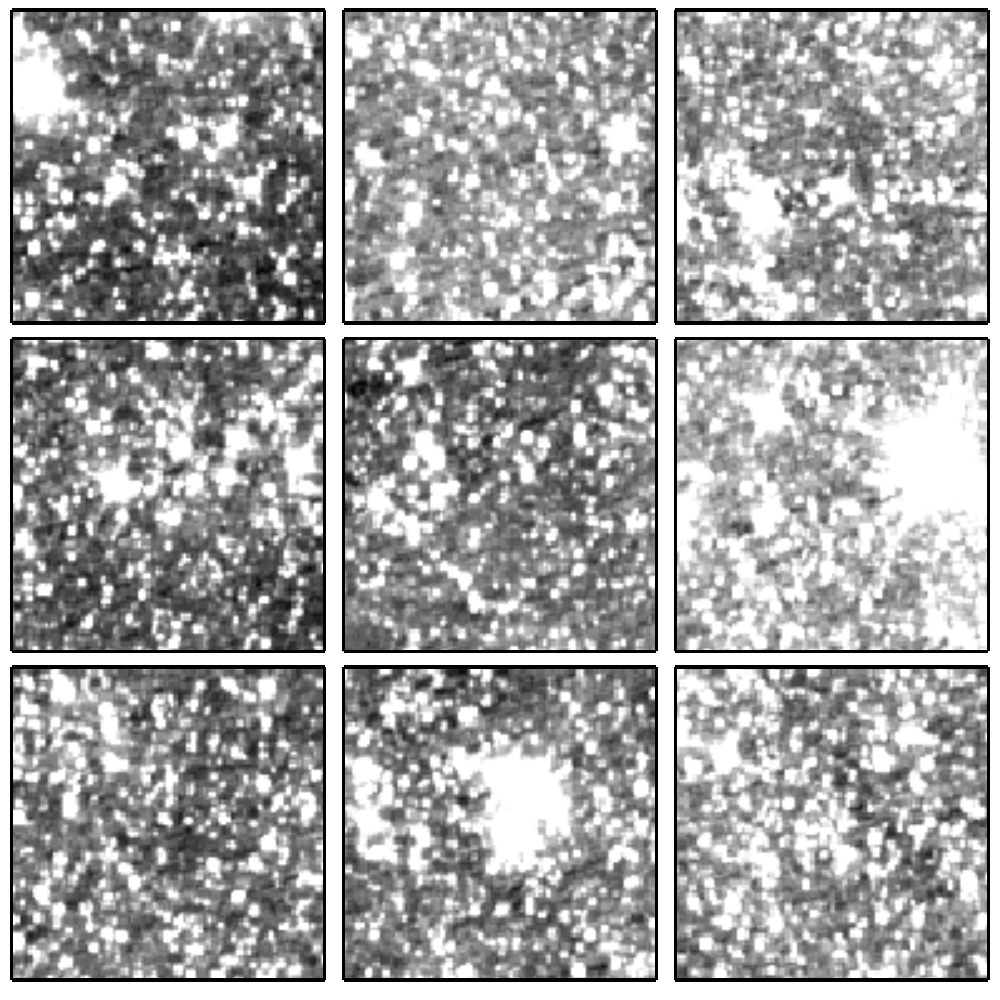} &
\includegraphics[width=0.4\textwidth]{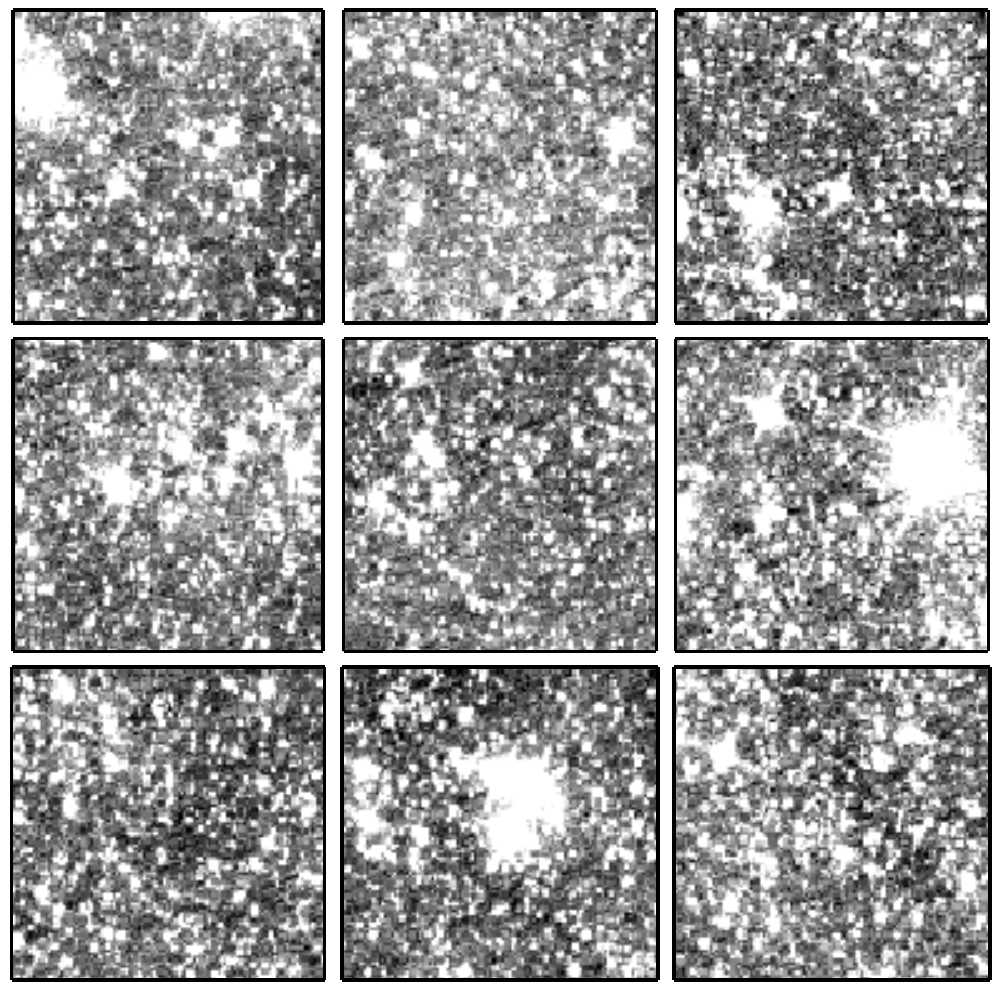}
\end{tabular}
\end{center}
\caption{Zoom-ins of corners of the W1 coadd tile 1384p454, to show
  variations in the background levels.  These variations would explain
  the broader-than-expected distribution of pixel values relative to
  the uncertainty estimates seen in the WISE coadds (see
  \figref{fig:pixdist}).  My unWISE coadds seem to have considerably less
  variation.
  \label{fig:bglevel}}
\end{figure}

\Figref{fig:pixdist} shows the distribution of pixel values with
respect to their error estimates.  I find that the AllWISE Release
coadd uncertainty values appear to underestimate the variation in
pixel values, at least at full-tile scales: The distribution of pixel
values in full Atlas Image tiles is considerably broader than expected
for a flat background.  This could be due to spatial variations in the
background, but since the same variations are not seen in my coadds,
I expect this could be due to variations in the AllWISE Release
background \emph{estimates}, rather than true variations in the
background levels.  The method involves matching the background of an
incoming frame with the existing coadd; thus ``the final background
level in a co-add is dictated by the first frame that was
[re]projected.''\footnote{WISE All-Sky Data Release Explanatory
  Supplement, section 4.4.f.iii.}  An example is shown in
\figref{fig:bglevel}.

\begin{figure}
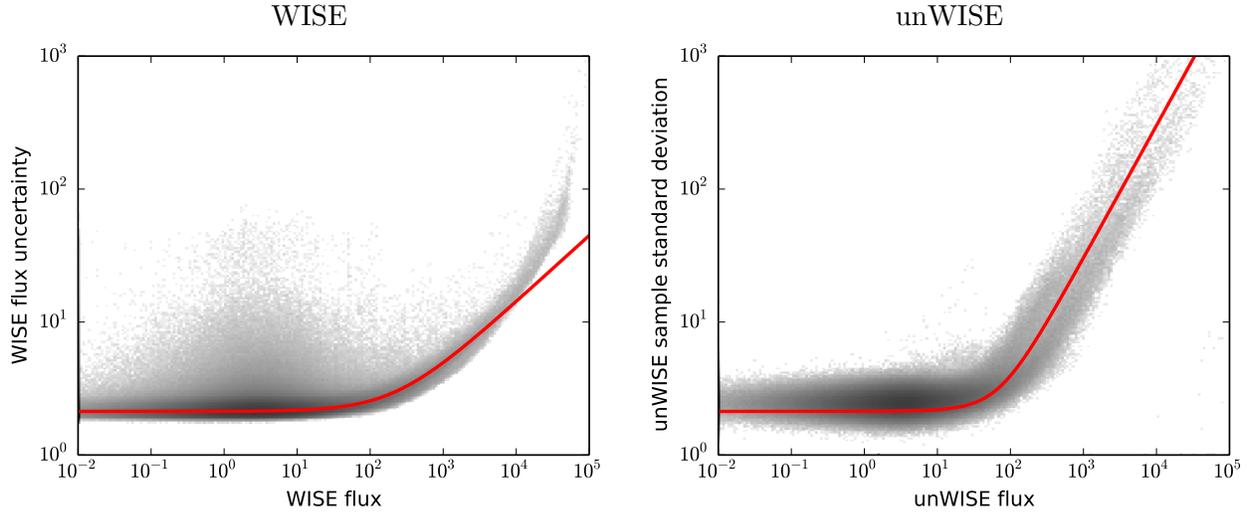

\begin{center}
\begin{tabular}{@{}cc@{}}
WISE & unWISE \\
\includegraphics[width=0.49\textwidth]{\bwfig{plots2/co-10}} &
\includegraphics[width=0.49\textwidth]{\bwfig{plots2/co-11}}
\end{tabular}
\end{center}
\caption{Pixelwise distributions of intensity (flux) and error
  estimates, for W1 tile 1384p454.  \emph{Left:} WISE \emph{intensity} and
  \emph{uncertainty} maps.  Note the log-log scales; the fluxes are scaled 
  to zeropoint 22.5; ie, these are ``Vega nanomaggies''.
  At low flux levels, the error is dominated by read noise and
  background Poisson noise.  At high flux level, the Poisson
  contribution from the sources becomes significant.  The AllWISE
  Release uncertainty estimate comes from propagating the error
  estimates from the input frames forward through the coaddition
  process, and does not include any systematic uncertainties.
  \emph{Right:} unWISE \emph{intensity} and \emph{per-pixel standard
    deviation} estimates, on the same axes.  This error estimate is
  based on the sample variance computed during coaddition.  Since it
  is based on the actual data values encountered, it includes
  systematic effects that produce scatter (but not bias).  The red
  curves are rough by-eye ``fits'' of the expected behaviors.  The
  WISE curve includes a constant noise floor plus Poisson noise whose
  standard deviation goes as the square root of flux.  The unWISE
  curve includes those same components plus a component whose standard
  deviation goes as the flux.  This would be expected if there were
  \emph{multiplicative} systematic errors; for example, if there were
  scatter in the input frame zeropoints or flat-fields.  The
  (fit-by-eye) curve plotted here includes a 3\% multiplicative error
  contribution.  I suspect the AllWISE Release coadds include
  multiplicative systematics of the same magnitude.
  \label{fig:pixdist2}}
\end{figure}

\Figref{fig:pixdist2} examines the behavior of my per-pixel sample
standard deviation measurements.  This can be seen as a measurement of
the uncertainty of the coadd based on the data values encountered
during coadding.  Interestingly, there are considerable differences
between this measurement of uncertainty and the ``error propagation''
approach used in the AllWISE Release; this could be explained by
multiplicative systematics of about 3\%.  I suspect the AllWISE
Release Atlas Images are also affected by this systematic scatter.

\begin{figure}
\setlength{\figh}{0.18\textwidth}
\begin{center}
\begin{tabular}{@{}r@{}c@{\hspace{1em}}c@{\hspace{1em}}c@{}}
&
{\small WISE} & 
{\small unWISE} &
{\small unWISE} \\
& &
{\small \makebox[0pt][c]{(no median filter)}} &
{\small \makebox[0pt][c]{(with median filter)}} \\
\raisebox{0.5\figh}{W3} &
\includegraphics[height=\figh]{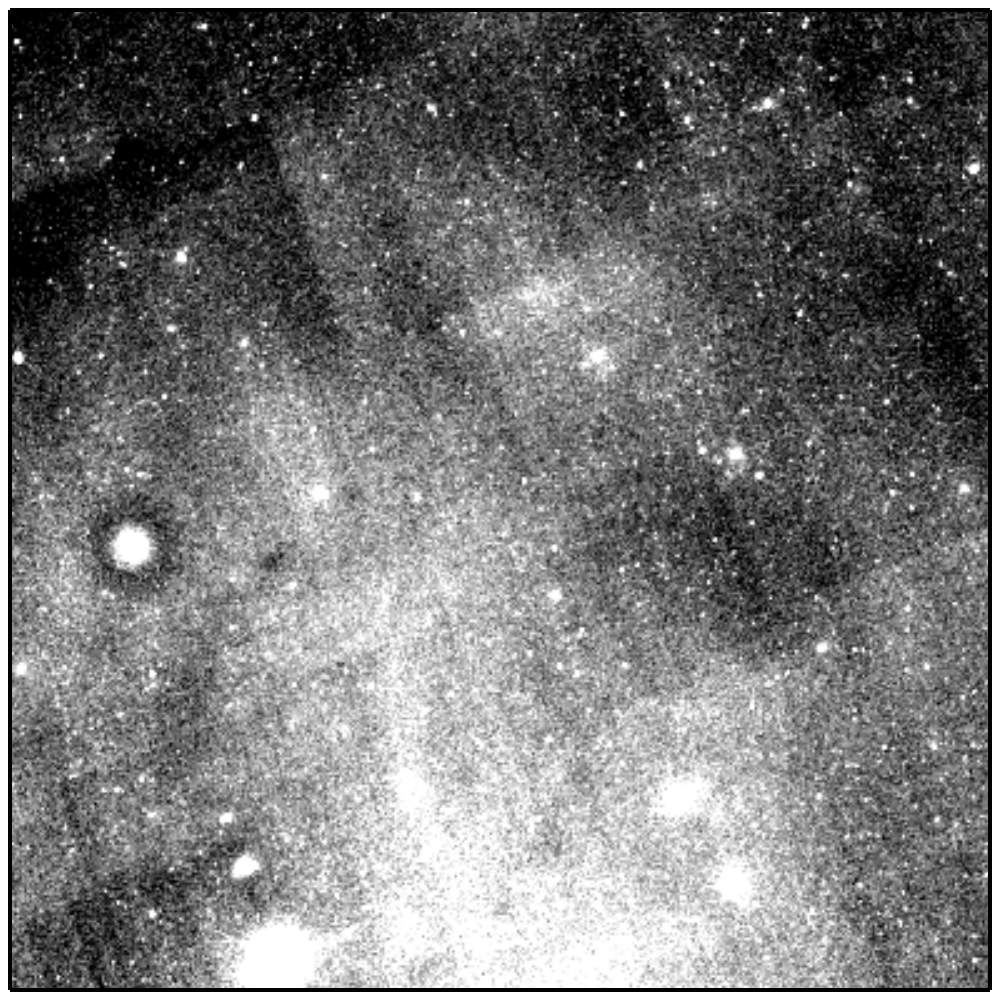} &
\includegraphics[height=\figh]{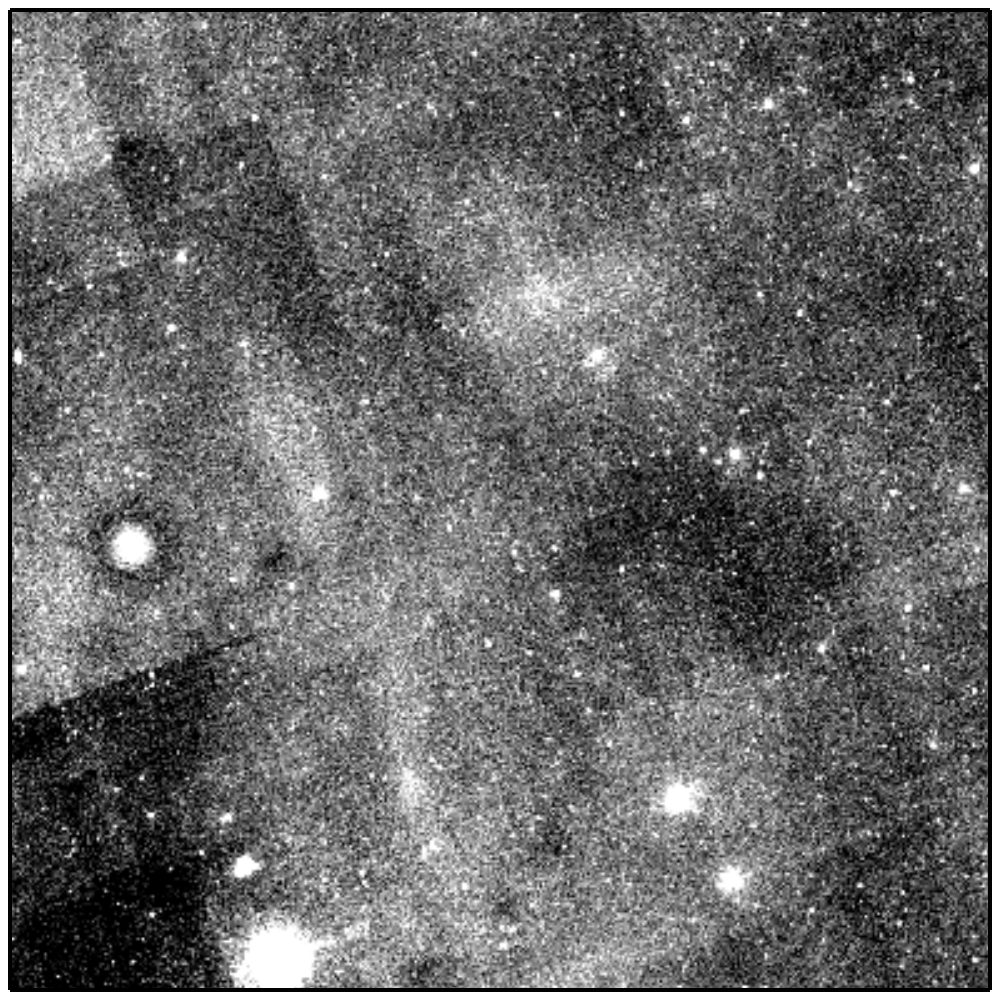} &
\includegraphics[height=\figh]{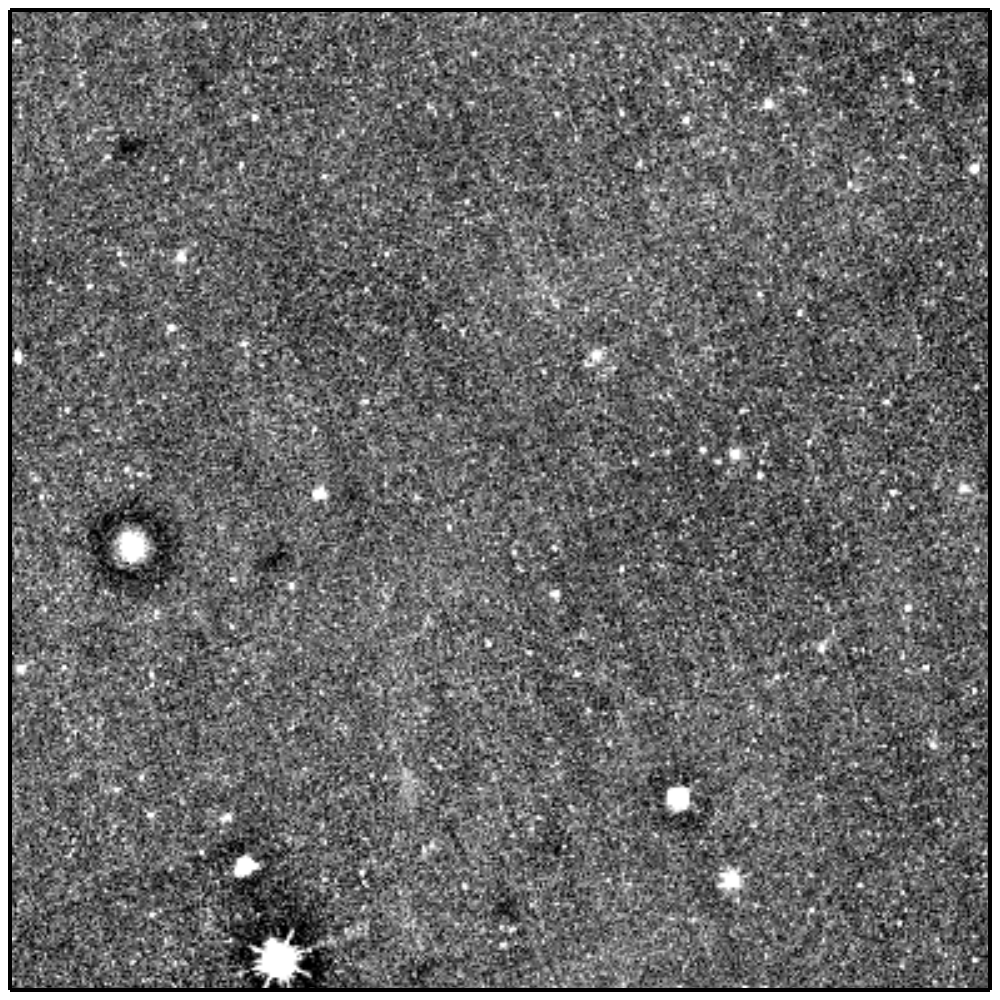} \\
\raisebox{0.5\figh}{W4} &
\includegraphics[height=\figh]{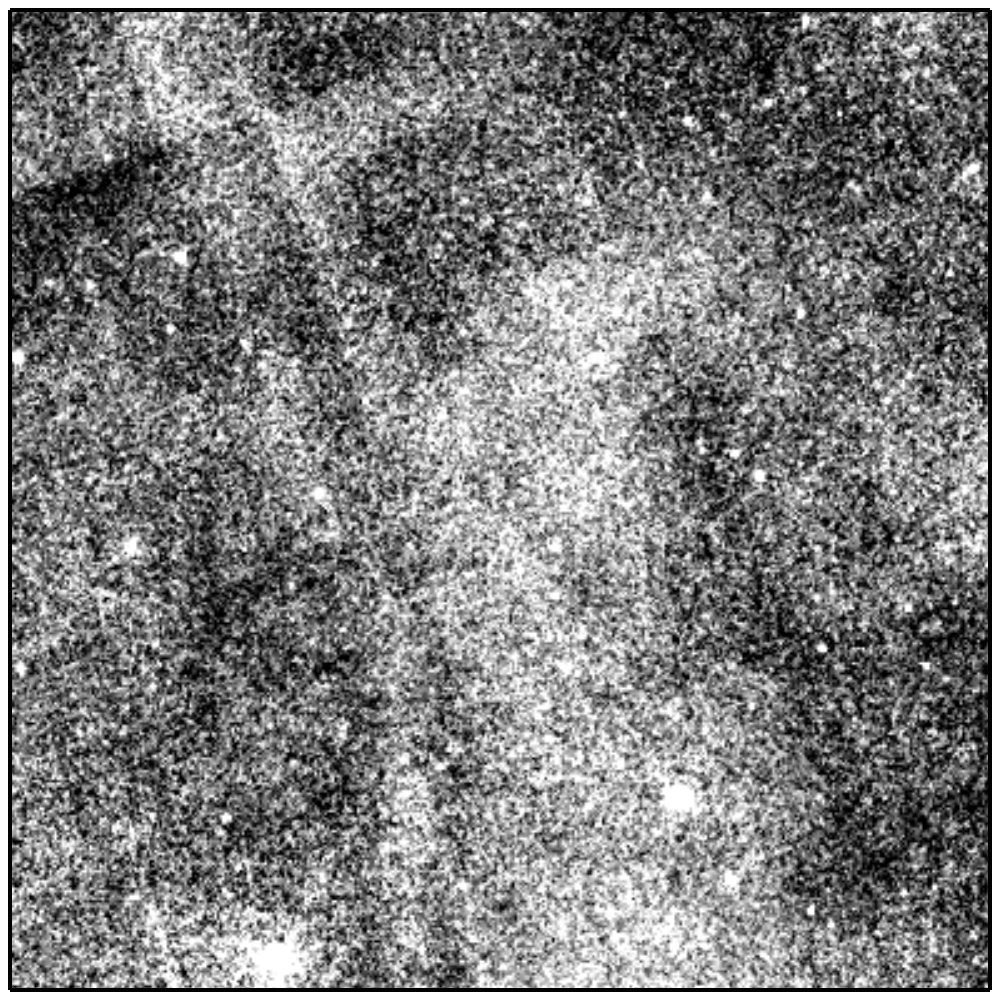} &
\includegraphics[height=\figh]{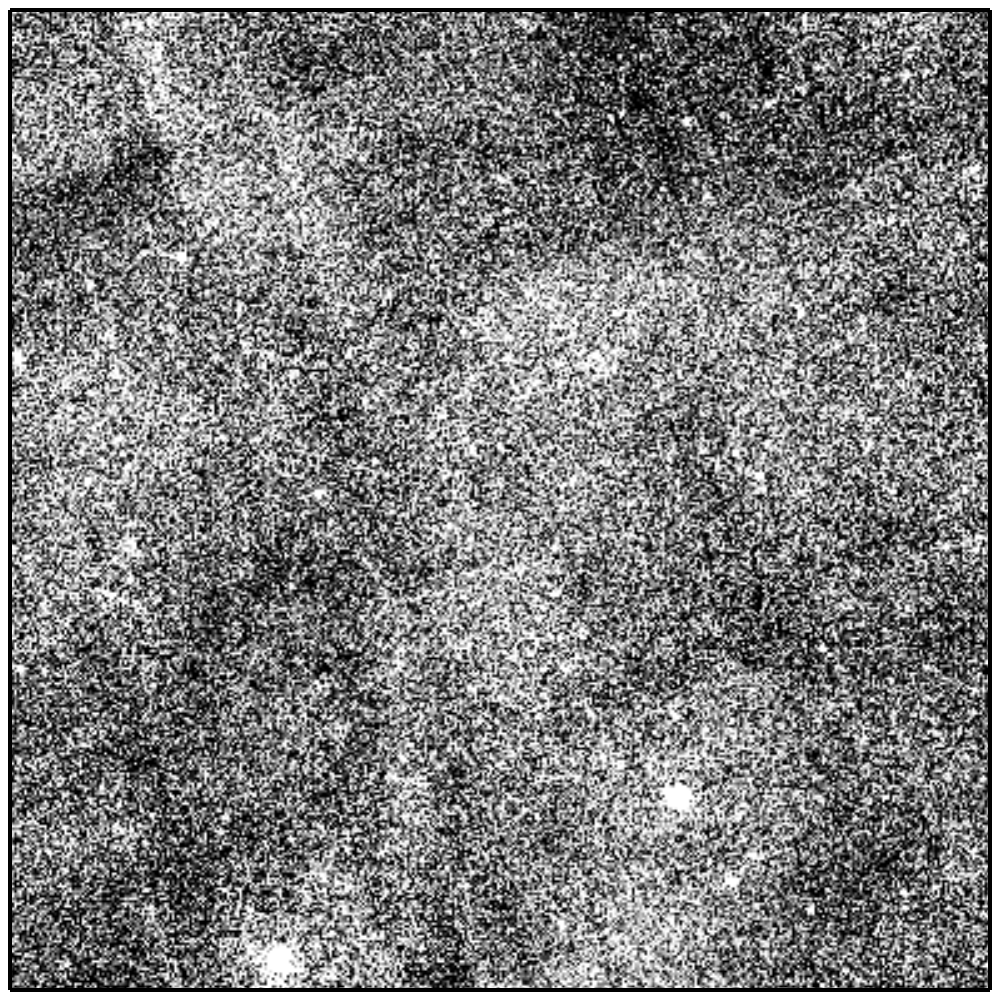} &
\includegraphics[height=\figh]{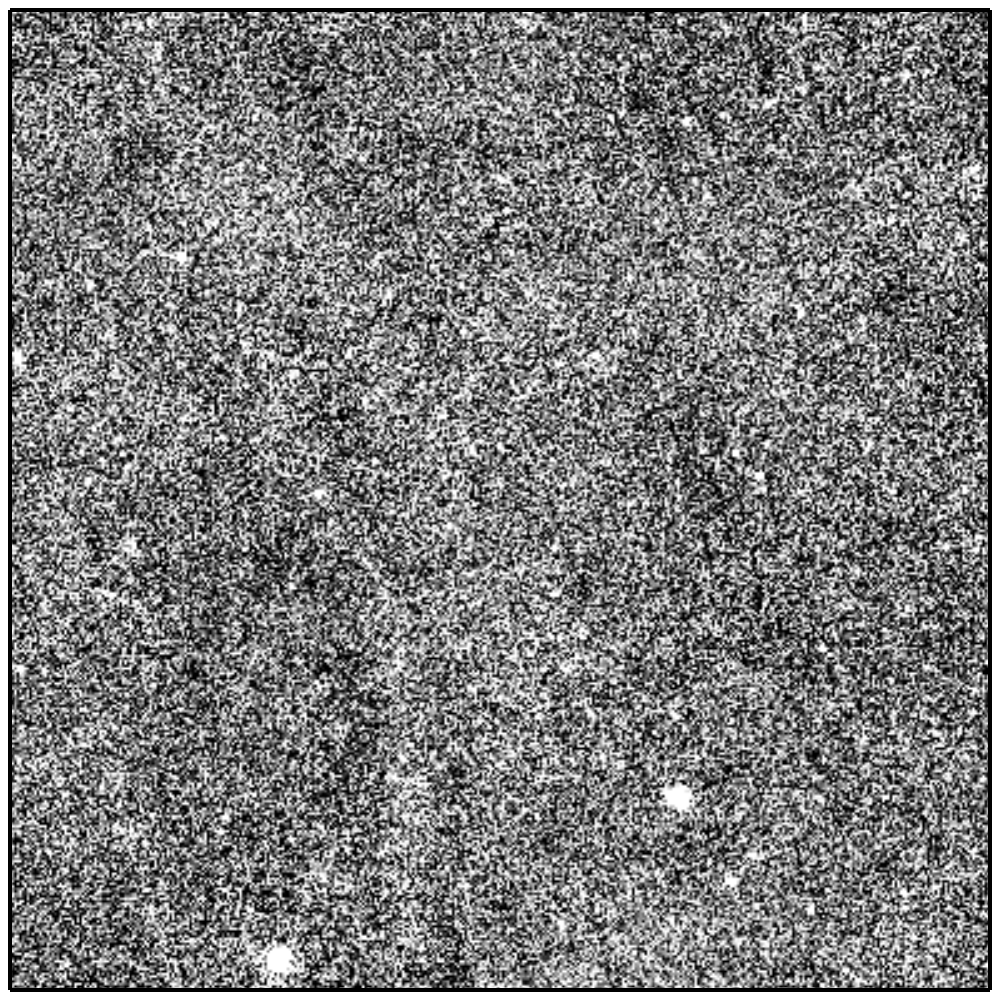}
\end{tabular}
\end{center}
\vspace{-1.5em}
\caption{Effect of median-filtering input frames in W3 and W4.
  \emph{Top row:} W3 images for tile 1384p454.  The AllWISE Release
  Atlas Image (left) shows considerable structure, some of which
  clearly follows the input frame boundaries.  I find visually similar
  artifacts if I omit the median filtering step (middle).  Applying a
  median filter (right), the spatially varying background is
  essentially gone.  Note the dark ring (suppressed background) around
  the bright star at the bottom of the frame.  (The bright ``star'' at
  the middle-left is a persistence artifact from this bright source.)
  \emph{Bottom row:} W4 images.
\label{fig:medfilt}}
\end{figure}

\begin{figure}
%\newlength{\figw}
\setlength{\figw}{0.24\textwidth}
\newcommand{\spc}{\hspace{0.01\textwidth}}
\begin{center}
\begin{tabular}{@{}p{\figw}@{\spc}p{\figw}@{\spc}p{\figw}@{\spc}p{\figw}@{}}
{\small WISE} & 
{\small unWISE} &
{\small unWISE} &
{\small unWISE (no m.f.,} \\
&
{\small (median filter)} &
{\small (no median filter)} &
%(no median filter, \newline 
{\small \mbox{background matched)}} \\
\includegraphics[width=0.24\textwidth]{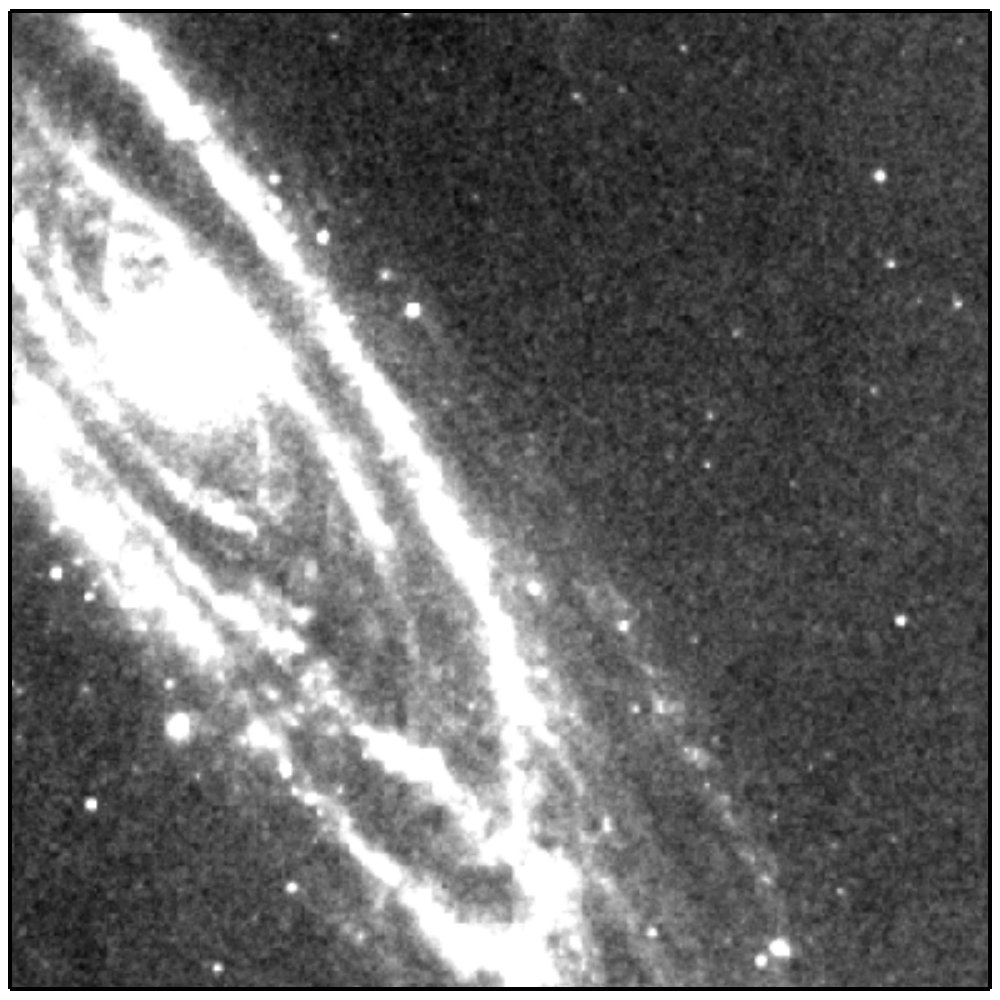} &
\includegraphics[width=0.24\textwidth]{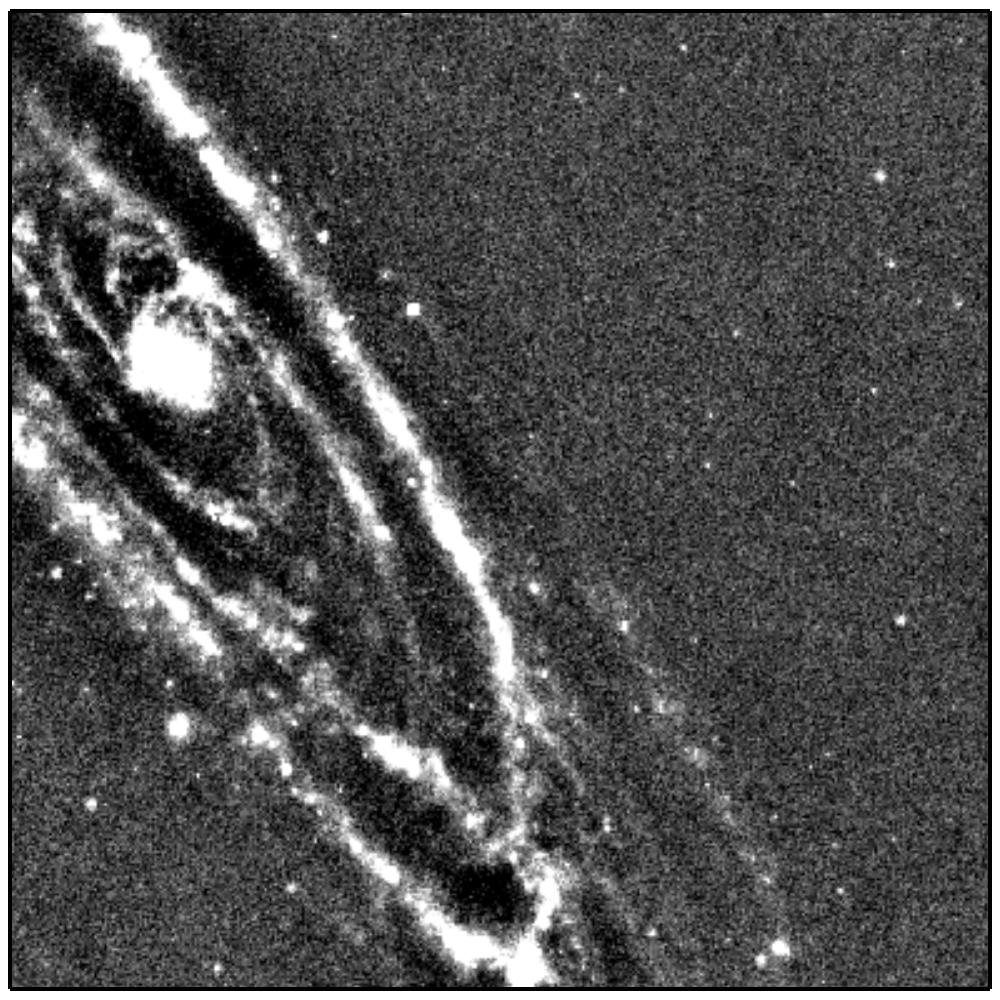} &
\includegraphics[width=0.24\textwidth]{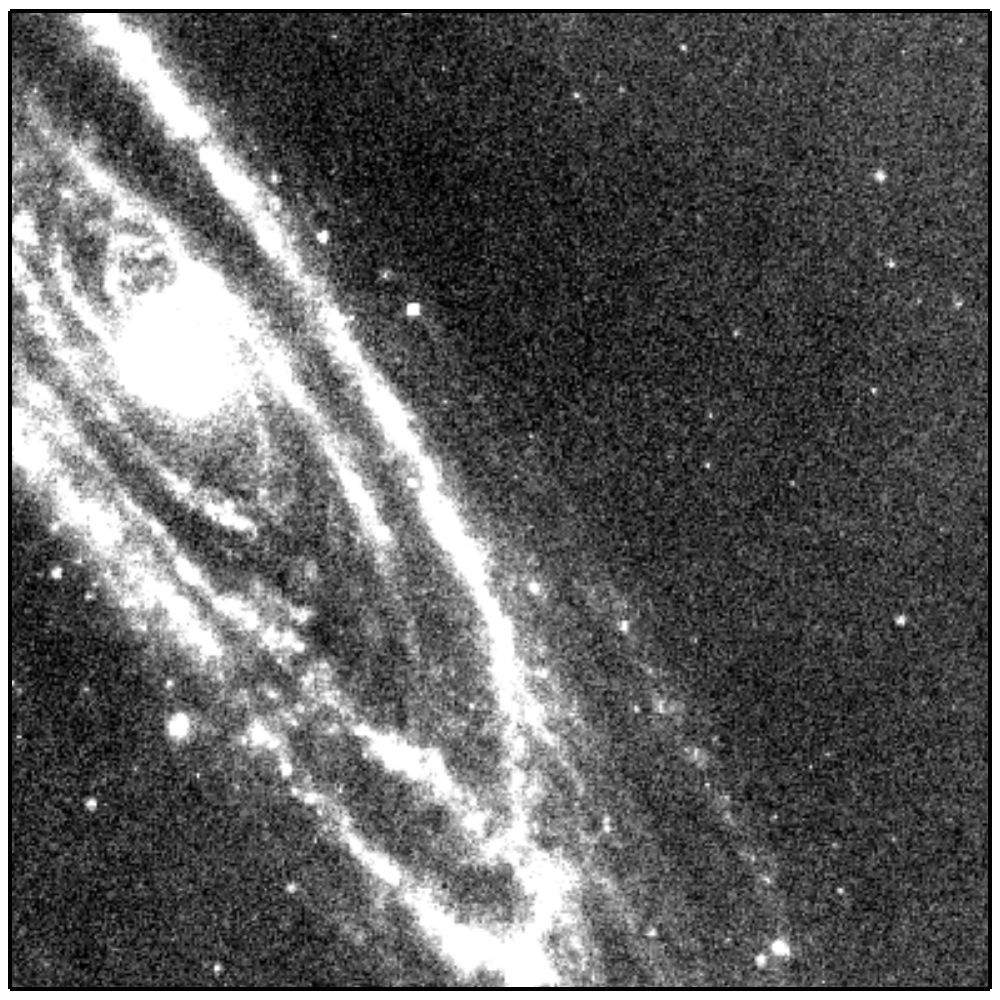} &
\includegraphics[width=0.24\textwidth]{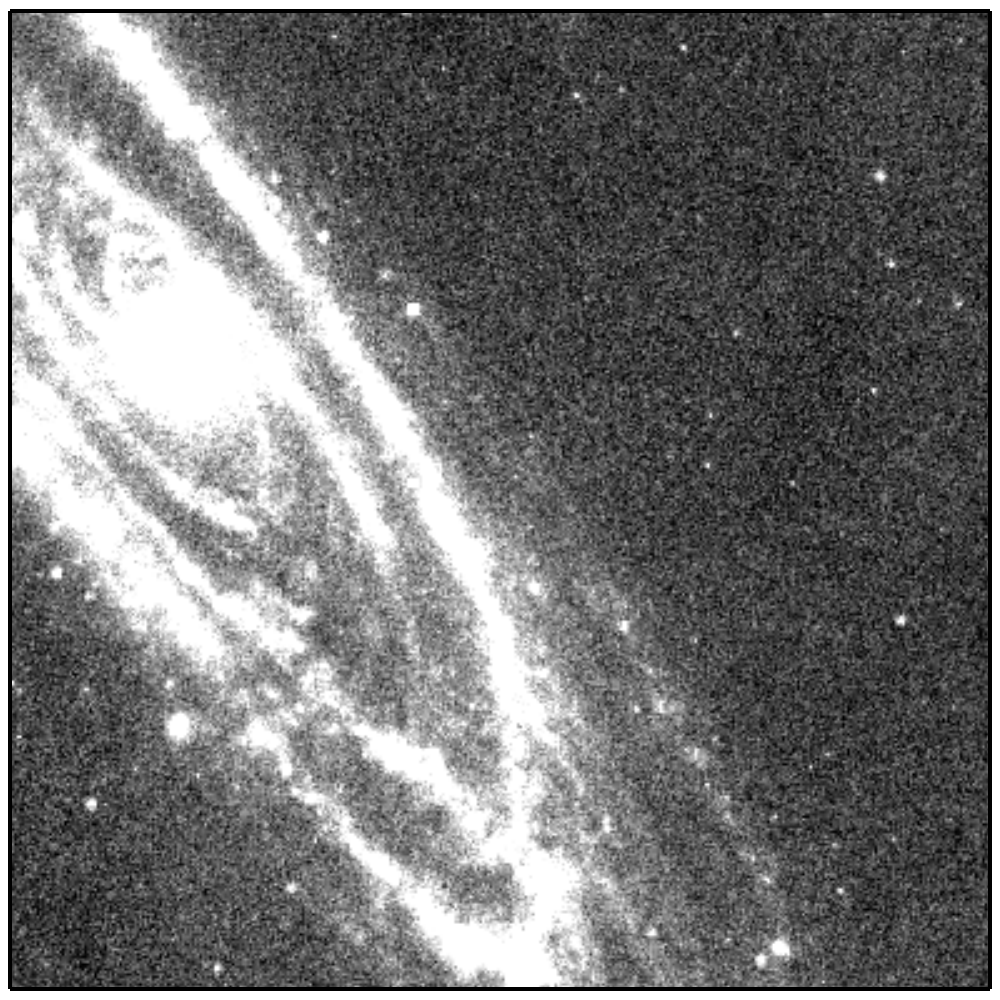}
\end{tabular}
\end{center}
\vspace{-1.5em}
\caption{Effect of median filtering in the presence of large, bright
  structures.  This is part of tile 0098p408, containing M31.
  \emph{Left:} The AllWISE image seems to show a slight depression in
  the background level above and right of the galaxy core.
  \emph{Second column:} My unWISE median-filtered image has a
  severely depressed background level near the galaxy core, since the
  local median within the galaxy is very high.  \emph{Third column:}
  with median-filtering disabled, I get results similar to AllWISE.
  \emph{Right:} with median-filtering off and background-matching on,
  the background looks smooth.
\label{fig:badmedfilt}}
\end{figure}

\Figref{fig:medfilt} demonstrates the effect of applying a spatial
median filter to estimate and remove a varying background from the
input frames before coadding them.  
As expected, the median filter removes large-scale structure from the
coadds, but also depresses the background levels around bright
sources.  A more extreme case of this is shown in
\figref{fig:badmedfilt}.
%

% 12 Gpixels included in coadd (imstat npole2/unwise-2709p666-w1-n.fits)
% Wall: 20872.46 s, CPU: 19292.66 s, VmPeak: 101720 MB, VmSize: 101432 MB, VmRSS: 96895 MB, VmData: 101113 M
% (but that was multi-threading the first-round)

As outlined above, I perform two rounds of coadding, where the first
round is used to reduce the impact of artifacts.  In my
implementation, I perform the resampling once, keeping the
resampled input frames in memory until the second round.  The memory
required thus increases with the number of pixels touching the coadd
tile.  In practice, this has not proven to be problematic: the tile
containing the north ecliptic pole (2709p666) has nearly $5\thou$ W1
exposures at its peak and is touched by approximately $10^{10}$
input-frame pixels in $20\thou$ exposures.  Producing this coadd tile
required peak memory of about $100$ Gbytes and took about $12$ hours
on a single CPU core.  The tile shown in \figref{fig:res1}, 1384p454,
is more typical: it is covered by $432$ exposures in W1, incorporates
a total of $\sim 3 \times 10^{8}$ input pixels, and took peak memory
of $2.3$ GB and $630$ seconds on a single CPU core.

%
% These plots were produced by the check-coadd.py script
% (paper_plots() function)
% at revs 23749/23750
%

%
% These plots were made via:
%
%  python -u unwise-coadd.py --force --outdir e --threads1 8 3000 > e3000.log  2>&1 &
%  python -u unwise-coadd.py --force --outdir f --threads1 8 --medfilt 50 3000 > f3000.log  2>&1 &
%  python -u unwise-coadd.py --force --outdir e --threads1 8 4000 > e4000.log  2>&1 &
%  python -u unwise-coadd.py --force --outdir f --threads1 8 --medfilt 50 4000 > f4000.log  2>&1 &
%  python check-coadd.py
% 
% At 23704
% (later, plots at 23797)
%

\section{Discussion}

The method presented here includes a number of heuristics to handle
the imperfections in real data.  These heuristics include noise
estimation for image weighting; masked-pixel patching; background
estimation; and masking of transients and defects in the imaging.

I weight each image in a coadd based on an estimate of an
image-wide (scalar) noise estimate.  I did not want to use pixel-wise
error estimates (which incorporate a Poisson noise term from the
source as well as the sky) since these lead to biased estimates
(pixels with fewer counts get a smaller error estimate and hence more
weight).  Using an error estimate based only on the baseline
background and read-noise should result in weights that optimize the
detectability of faint (background-dominated) isolated point sources.
I chose a simple heuristic for estimating this scalar error estimate:
I take the median of the unmasked uncertainty pixels from the WISE
single-frame pipeline.  Since these uncertainty pixels include a
component due to Poisson noise from the signal as well as read noise,
the median value is slightly biased high.  I could instead attempt to
estimate the mode of the uncertainty distribution in the same way I
estimate the background level.

Since I resample the input frames using the Lanczos kernel, masked
pixels in the input images would lead to large holes (as large as the
support of the kernel) in the resampled images.  Since several percent
of the pixels in typical WISE images are masked, this would have
resulted in unacceptably large losses, so I chose to ``patch''
pixels.  My crude approach is simply to average neighboring valid
pixels.  Other approaches for patching defects include the linear
predictive filter used by the Sloan Digital Sky Survey
\citep{sdss-edr}.  Patching pixels can be seen as redistributing the
weights of the interpolation kernel from the masked pixel to its neighbors.
I mask the resampled pixel closest to the patched pixel, so it is not
included in the ``masked'' coadds, $\coadd_m$.  Nearby resampled
pixels also include some contribution from the patched pixel; I chose
to include these since their effect should be rather small.
A more principled (but much more expensive) approach to patching
images would be to fit a forward model to a small image patch and take
the model prediction (with the model fit to unmasked pixels) as the
patched pixel value.  This would take into account the point-spread
function, nearby sources, and many nearby pixels, resulting in
predictions that were consistent with the modelled astrophysical
scene.

For the W1 and W2 bands, I assume flat backgrounds and estimate a
single scalar background level.  For W3 and W4, I found it helpful to
fit a varying background, and I chose to use a spatial median filter
for this purpose.  If I had a better model for the backgrounds (from
astrophysical and instrumental sources) and could fit or marginalize
over this model, that would comes closer to producing ideal coadds.  A
median filter should generally remove trends on spatial scales of
order the filter size and preserve features smaller than the filter,
but will introduce biases near bright sources and crowded
regions; see \figref{fig:badmedfilt} for example.  It would perhaps be
better to avoid including in the median estimate any pixels that
appear to be ``contaminated'' by sources (for example, by skipping
pixels that are significantly different than their neighbors), though
this is just a further heuristic rather than a real solution to the
problem of estimating backgrounds.  I am currently exploring possible
approaches to reducing the highly depressed background level around
bright structures (as shown in \figref{fig:badmedfilt}).  One option
would be to also median-smooth on a larger spatial scale and add those
filtered values back in, though this would likely still result in
artifacts around structures in a range of scales.  Another option
would be to find the median or some other statistic of the
\emph{stack} of median-smoothed values, adding this back to the coadd.
This way, extended structures that are \emph{consistent} in many
individual exposures would appear in the coadd.  Yet another approach
would be to match the large-scale variations to those found in other
data sets, as done by \cite{meisner}; indeed, their maps, with point
sources and other small-scale structures removed, could be ideal for
repairing the large-scale structures damaged by my median filtering.

The AllWISE Atlas Images were created using a ``background-matching''
approach for removing frame-to-frame variations in the background
level.  In this approach, as a frame is being accumulated into the
coadd, its median value is matched to that of the existing coadd, in
the region where they overlap.  This approach tends to produce
backgrounds that are smooth on large scales (see
\figref{fig:badmedfilt}), even in the presence of bright objects,
since it uses no explicit estimation of the background level.
However, it has the disadvantages that it is order-dependent---the
first few frames accumulated into the coadd have disproportionately
large influence on the final background---and also depends on the
extent of the coadd tile: shifting the tile boundaries results in
different backgrounds.

I attempt to mask transients and defects in the individual exposures
by building an initial coadd to estimate the mean and variance of the
input frame pixels contributing to each coadd pixel.  This approach of
storing only summary statistics is in contrast to the approach used in
the AllWISE Release, of keeping in memory the full list of input frame
pixels contributing to each coadd pixel, and then performing
operations on those pixel lists.  Using robust statistical estimates,
the WISE team's approach should allow nearly 50\% of the input frames
to be contaminated without affecting the results.  This power
presumably comes at significant computational cost and complexity.
My simpler and cheaper approach of accumulating only the mean and
variance (and removing an input frame's contribution to these
estimates before evaluating that frame) should be expected to work
well at flagging a single moderate outliers in each pixel, or multiple
extreme outliers if the number of contributing frames is large.  If
there is more than a single outlier, the sample variance estimate will
be inflated by the outlier, and moderate outliers will no longer
deviate from the mean enough to be flagged.  Despite these expected
shortcomings, I have found that my simple approach seems to work
well in practice.  Possible improvements would be to detect pixels
whose sample variance is larger than would be expected from a Poisson
noise model, and do more careful outlier detection there.

In attempting to remove artifacts such as satellite trails and cosmic
rays, we also remove fast-moving sources and clip sources that vary in
brightness.  One would expect variable sources to induce a larger
sample variances than a non-varying source of the same mean magnitude,
and this might be a simple way of flagging such sources.  In general,
though, these coadds should be used for the non-transient sky only.

One might ask why I felt it necessary to write a new code to produce
these coadds when publicly available codes such as SWarp \citep{swarp}
already exist.  The core coaddition (image warping and interpolation)
could indeed have been implemented using SWarp, but the detection and
masking of transients and defects (which I do by creating first-round
coadds) would have required first running SWarp to do the
interpolation and first-round coadds, then detecting and masking
transients in the ``temporary'' files written by SWarp, then running
the second-round coadds on these altered files.  I felt it would be
simpler to reimplement the warping and interpolation functionality.

% The code developed to produce these coadds has some WISE-specific
% parts (for example, the use of metadata from the WISE pipeline for
% frame selection and moon-affected frame elimination), and makes some
% assumptions about the data---that the point-spread function is
% approximately constant; that they are well-sampled with roughly
% uniform noise; that the

The WISE team should be commended on their timely, high-quality data
releases.  One shortcoming I have attempted to address in this work is
that the Atlas Images are blurred by convolution by the point-spread
function.  I present a new set of coadded images that preserve the
resolution of the original exposures.  I use a theoretically justified
resampling method with some approximations.  In the process of
validating my results, I found that my pixelwise sample variances, and
likely the AllWISE Release Atlas Images as well, include a $\sim$3\%
systematic uncertainty in addition to the expected statistical
uncertainty.  These images, and the code used to produce them, are
publicly available at \niceurl{http://unwise.me}.

% This work was only possible due to the availability of
% well-calibrated, standards-compliant ``level 1b'' images.

\acknowledgements
It is a pleasure to thank
  David W.~Hogg (NYU),
  David Schlegel (LBL),
  Tod Lauer (NOAO),
  Aaron Meisner (CfA), and
  the anonymous referee
for helpful comments on the manuscript;
  Rachel Mandelbaum (CMU)
for enlightening discussions;
  David Schlegel,
  Stephen Bailey,
  and Peter Nugent (LBL)
for data-management and computing assistance;
  Michael Blanton (NYU) for the spatial median-filtering code;
and Roc Cutri (IPAC) and the rest of the WISE team for very helpful
responses to numerous queries at the WISE Help Desk.

This publication makes use of data products from the Wide-field
Infrared Survey Explorer, which is a joint project of the University
of California, Los Angeles, and the Jet Propulsion
Laboratory/California Institute of Technology, and NEOWISE, which is a
project of the Jet Propulsion Laboratory/California Institute of
Technology. WISE and NEOWISE are funded by the National Aeronautics
and Space Administration.

This research used resources of the National Energy Research
Scientific Computing Center (NERSC), which is supported by the Office
of Science of the U.S. Department of Energy under Contract
No.~DE-AC02-05CH11231.

This research has made use of NASA's Astrophysics Data System.

% \textit{Facilities:} \facility{WISE}.

\end{document}